\documentstyle[12pt]{article}

\newlength{\dinwidth}
\newlength{\dinmargin}
\def\lapproxeq{\lower .7ex\hbox{$\;\stackrel{\textstyle
<}{\sim}\;$}}
\def\gapproxeq{\lower .7ex\hbox{$\;\stackrel{\textstyle
>}{\sim}\;$}}
\def\funp{{I\!\!P}}
\def\be{\begin{equation}}
\def\ee{\end{equation}}
\def\bea{\begin{eqnarray}}
\def\eea{\end{eqnarray}}

\begin{document}
\titlepage
\begin{flushright}
DTP/97/18\\
March 1997 \\
\end{flushright}

\begin{center}
\vspace*{2cm}
{\Large \bf A unified BFKL and GLAP description of $F_2$ data }
\\
\vspace*{1cm}
J.\ Kwiecinski$^{a,b}$, A.D.\ Martin$^a$ and A.M.\ Stasto$^{a,b}$

\end{center}
\vspace*{0.5cm}
\begin{tabbing}
$^a$xxxx \= \kill
\indent $^a$ \> Department of Physics, University of Durham,
Durham, 
DH1 3LE, UK. \\

\indent $^b$ \> H.\ Niewodniczanski Institute of Nuclear Physics,

Department of Theoretical Physics, \\
\> ul.\ Radzikowskiego 152, 31-342 Krakow, Poland.
\end{tabbing}

\vspace*{2cm}

\begin{abstract}
We argue that the use of the universal {\it unintegrated} gluon
distribution and
the $k_T$ (or high energy) factorization theorem provides the
natural framework for
describing observables at small $x$.  We introduce a coupled pair
of evolution equations for the unintegrated 
gluon distribution and the sea quark distribution which
incorporate both the
resummed leading $\ln (1/x)$ BFKL contributions and the resummed
leading
$\ln (Q^2)$ GLAP contributions.  We solve these unified equations
in the
perturbative QCD domain using simple parametic forms of the
nonperturbative 
part of the {\it integrated} 
distributions.  With only two (physically 
motivated) input parameters we find that this $k_T$ factorization
approach gives an 
excellent description of the measurements of $F_2 (x,Q^2)$ at
HERA.  In this way the 
unified evolution equations allow us to determine the gluon and
sea quark 
distributions and, moreover, to see the $x$ domain where the
resummed 
$\ln (1/x)$ effects become significant.  We use $k_T$
factorization to predict
the longitudinal structure function $F_L (x,Q^2)$ and the charm
component of
$F_2 (x,Q^2)$.
\end{abstract}

\newpage
\noindent {\large \bf 1.  Introduction}

The experiments at HERA have opened up the small Bjorken $x$
regime.  One of
the most striking features of the data is the strong rise of the
structure 
function $F_2$ as $x$ decreases from $10^{-2}$ to below $10^{-4}$
\cite{HERA}.  At
first sight it appeared that the rise was due to the (BFKL)
resummation of leading
$\ln (1/x)$ contributions \cite{BFKL,LIP}.  However, with an
appropriate choice of input 
distributions and of the starting scale for the $Q^2$ evolution,
the observed growth can also be reproduced within the 
conventional GLAP framework which just sums up the leading (and 
next-to-leading) $\ln Q^2$ contributions.  Indeed GLAP global
fits exist
which give a good description of the small $x$ measurements of
$F_2$ in the $x$ 
range accessible at HERA \cite{MRS,CTEQ}, see also
\cite{GRV,SPAN}. Is it possible 
to conclude that there will be no significant BFKL-type
contributions to $F_2$ in the HERA 
small $x$ regime or does a physically reasonable alternative
description exist 
with sizeable $\ln(1/x)$ resummation contributions? Here we
address this 
question. Clearly the specification of the non-perturbative input
to the 
QCD evolution will be crucial.

Recall that the basic dynamical quantity at low $x$ is the gluon
distribution
$f(x,k_T^2)$ unintegrated over its transverse momentum $k_T$. It
is related to
the conventional gluon density $g(x,Q^2)$ by
\be
xg(x,Q^2) \; = \; \int^{Q^2} \: \frac{dk_T^2}{k_T^2} f(x,k_T^2).
\label{eq:z1}
\ee
In the leading $\ln (1/x)$ approximation $f(x,k_T^2)$ satisfies
the BFKL 
equation and exhibits an $x^{-\omega_0}$ growth and a diffusion
in $\ln k_T^2$ as 
$x \rightarrow 0$, where the BFKL or so-called hard Pomeron
intercept $\omega_0 
= (3 \alpha_S/\pi) 4 \ln 2$ for fixed
$\alpha_S$.  The observable quantities are computed in terms of
$f$ via the $k_T$ (or high energy) 
factorization prescription \cite{KTFAC,CIAFKT}.  For example, the
structure functions $F_i$ are 
given by
\be
F_i \; = \; F_i^{\gamma g} \: \otimes \: f
\label{eq:z2}
\ee
where $\otimes$ denotes a convolution in transverse, as well as
longitudinal, 
momentum. Here $F_i^{\gamma g}$ are the off-shell gluon structure
functions, 
which at lowest order are determined by the quark box (and
crossed-box)
contributions to photon-gluon fusion, see Fig.~1.

The BFKL gluon $f(x,k_T^2)$ and $k_T$ (or high energy)
factorization theorem 
were used \cite{AKMS1} to predict the small $x$ behaviour of
$F_2$ prior to the 
measurements at HERA.  The method used a starting distribution
$f(x_0,k_T^2)$ at, 
say, $x_0 \; = \; 0.01$ deduced from an integrated gluon
$g(x_0,Q^2)$ which had 
been determined in a global parton analysis of fixed-target deep
inelastic and related 
scattering data.  With this input the BFKL equation was solved to
determine $f(x,k_T^2)$ 
in the small $x$ domain $(x<x_0)$.  A major uncertainty in this
procedure to predict 
$F_2$ is the treatment of the infrared region, $k_T^2 < k_0^2$. 
A recent study 
\cite{BE} using this approach finds that the BFKL predictions for
$F_2$ increase 
too steeply with decreasing $x$ in comparison with the HERA
measurements, and
concludes that there is no evidence for the effects of the
resummation of
$\ln (1/x)$ terms.  Before we accept such a conclusion we should
note the 
limitations of this form of the test of the BFKL
$k_T$-factorization approach.

It is assumed \cite{BE,AKMS2} that in the infrared region  
$f(x,k_T^2)$ has the form
\be
f(x, k_T^2 < k_0^2) \; = \; \frac{k_T^2}{k_T^2 + k_a^2} \: 
\frac{k_0^2 + k_a^2}{k_0^2} \: f (x, k_0^2) 
\label{eq:z3}
\ee
with $k_0^2 = 1$ GeV$^2$ say, and where $k_a^2$ is an adjustable
parameter.  
It turns out that there is a sizeable contribution from the
infrared 
non-perturbative region which, since it is tied to $f(x,k_0^2)$,
is forced to 
have the BFKL growth with decreasing $x$.  The formalism does not
include the 
GLAP leading $\ln Q^2$ 
resummations which go beyond the leading $\ln (1/x)$
approximation. Finally it
should be noted that the BFKL equation for $f (x, k_T^2)$ only
resums the leading order
$\ln (1/x)$ terms.  Sub-leading $\ln (1/x)$ effects are expected
to reduce the
growth of $f$ with decreasing $x$.

Clearly this simplified procedure provides only a crude test of
the underlying
dynamics in the small $x$ (HERA) domain.  The main deficiencies
are the treatment
of the infrared region and the lack of a unified approach which
incorporates both
the BFKL $\ln (1/x)$ and the GLAP $\ln (Q^2)$ resummations. An
important
development is the demonstration by \cite{KTFAC,CIAFKT,AKMS1}
that at the 
leading twist level the BFKL $k_T$ factorization approach can be
reduced to the conventional collinear GLAP 
factorization in which the anomalous dimensions and coefficient
functions are
extended to include the full resummation of leading $\ln (1/x)$
terms.  This 
motivated an informative study by Thorne \cite{T}.  He has shown
(in a scheme 
independent way) that the inclusion of the $\ln (1/x)$ terms
within the collinear 
factorization approach gives a satisfactory, and even an
improved, description
of the $F_2$ data. He assumes non-perturbative components to the
input 
distributions for the observables at scale $Q_0^2$
which are \lq flat' at small $x$, and which are the 
{\it only} contributions 
at the scale $(A_{LL} \lapproxeq 1 {\rm GeV}^2)$ 
that denotes the boundary between the perturbative and
non-perturbative regions.  In this way he demands that the
rise of the structure functions with decreasing $x$ must entirely
come from 
perturbative effects.  In summary Thorne finds, within the
collinear factorization 
framework, that the \lq BFKL' $\ln (1/x)$ terms are not excluded,
but rather are 
favoured, by the $F_2$ data.

The agreement with the data obtained by Thorne \cite{T} is
contrary to the 
conclusion of ref.\ \cite{BE} 
which was based on the BFKL gluon and $k_T$ factorization.  This
suggests that
the application of the $k_T$ factorization approach was too
simplistic.  Here we re-examine and improve the determination of 
$f (x, k_T^2)$ and, via $k_T$ factorization, obtain a more
realistic description 
of $F_2$. 

The first improvement is that we study a \lq unified' equation
for $f (x, k_T^2)$
which incorporates BFKL and GLAP evolution on an equal footing
\cite{GLR}.  To be precise
we solve a coupled pair of integral equations for the gluon and
sea quark
distributions, as well as allowing for the effects of the valence
quarks.  In 
this way we eliminate the problems of matching at $x = x_0$.  A
second  
improvement is a more physical treatment of the non-perturbative
(or infrared)
contributions to the BFKL equation and the $k_T$ factorization
integrals.  We 
shall see that the former can be specified entirely by the {\it
integrated} gluon
distribution at the scale $Q^2 = k_0^2$ which marks the boundary
of the perturbative
and non-perturbative regions, whereas the integrals  also need a
non-perturbative
component of the sea. In fact we find that an excellent
description of the HERA 
measurements of $F_2$ is possible in terms of just two physically
motivated
parameters which fully determine these infrared contributions.

There is considerable merit in going back from the collinear to
the $k_T$ 
factorization approach for the description of small $x$ deep
inelastic scattering.  
Indeed, in the reduction to collinear form we loose some of the
physical 
structure which is contained in the gluon ladder and $k_T$
factorization. We 
discuss these limitations of the collinear approach below.

First we note that the high energy (or low $x$) behaviour of the
structure
functions is driven by the BFKL gluon ladder coupled (through
$k_T$ factorization) to the photon via the
quark box.  The corresponding Feynman diagram has a calculable
perturbative
contribution for all $Q^2$ (except for $Q^2 \rightarrow 0$ for
massless quarks).  
The \lq hard' or \lq QCD' pomeron contribution generated by this
ladder
is thus present for all $Q^2$.  In particular there is a known
perturbative
contribution to the structure functions at $Q^2 = k_0^2$ coming
from 
configurations in which the gluon transverse momenta within the
ladder lie in the 
perturbative domain $k_T^2 > k_0^2$.  This contrasts with the
collinear 
factorization approach in which it is contrived to describe the
observables in
terms of a purely non-perturbative contribution at some scale,
say $Q^2 = k_0^2$. 
The perturbative component, which must be present at $Q^2 =
k_0^2$, will be 
evident in the $k_T$ factorization approach that we introduce
below.

A more subtle limitation of the Renormalisation Group (RG)
and collinear factorization approach at small $x$ concerns the
treatment of the
running of $\alpha_S$.  We solve the BFKL equation with running,
rather than fixed,
coupling where $\alpha_S$ depends on the local scale $k_T^2$
along the ladder.  
This way of implementing the running of $\alpha_S$ is supported
by the calculation
of next-to-leading $\ln (1/x)$ effects \cite{CC1,CC2}.  The
solution of the BFKL
equation with running $\alpha_S$ can be reduced to the
conventional RG form using
saddle point techniques.  However, the saddle point approximation
is not applicable
for arbitrarily small values of $x$
\cite{JKJC,LIP,CC1,CC2,KM,LEVIN}.

A third advantage of the $k_T$ factorization approach is that it
allows us to 
appropriately constrain the transverse momenta of the emitted
gluons along the 
BFKL ladder.  We are therefore able to quantify the effect of
imposing this 
constraint.  (Recall that in the usual application of the BFKL
equation the gluon
transverse momenta are taken to be unlimited.)  We will see that
the kinematic 
constraint largely subsumes the angular ordering constraint which
is the basis 
of the \lq CCFM' equation \cite{CCFM,KMS}.  The CCFM equation
also incorporates
both BFKL and GLAP evolution.

Another difference is that the BFKL contribution is a sum over
all twists, whereas
when it is reduced to collinear form only the leading twist is
conventionally
retained.  Finally $k_T$ factorization is much simpler to
implement at small
$x$ than collinear factorization.  We deal with dynamical
quantities (namely the
BFKL kernel and the structure function of the off-shell gluon)
which can be 
calculated perturbatively.  We calculate them to first order in
$\alpha_S$.  
Essentially the $\alpha_S \ln (1/x)$ terms are effectively
resummed by simply
integrating over the entire $k_T^2$ phase space allowed for the
gluon ladder and
the $k_T$ factorization integrals.

In summary, the natural framework with which to describe
observables at small
$x$ is the {\it unintegrated} gluon density $f (x, k_T^2)$
together with the
$k_T$ {\it factorization theorem}.  (Here we use it to calculate
the observable
deep inelastic scattering structure functions $F_2$ and $F_L$.)
That is, at small $x$ the distribution 
$f (x, k_T^2)$ is the basic, universal quantity which can be
taken from process
to process.  If we were to reduce this framework to a collinear
factorized form
then the \lq \lq simple", but rich, physics structure of the
gluon ladder is not
fully taken into account and even may be distorted.  It could be
argued that most
of the effects occur at subleading order in $\ln (1/x)$ and since
only the
leading order is completely known, little is lost.  However, the
effects have a
direct physical origin.  They are clearly present and are
expected to be the
dominant corrections in a more complete analysis.

Since we shall unify the BFKL and GLAP formalisms, the resulting
equation for
$f (x, k_T^2)$ is valid both for small
$x$ and large $x$. Moreover the region where the \lq BFKL' $\ln
(1/x)$ effects become 
significant will be decided by the underlying dynamics (QCD). We
use the formalism
to fit to deep inelastic data and so we are able to quantify the
importance of 
BFKL effects.

The outline of the paper is as follows.  In Section 2 we make
modifications to 
the BFKL equation for the unintegrated gluon distribution $f (x,
k^2)$ which allow for the
GLAP leading $\ln Q^2$ contributions and which enable all the
non-perturbative
effects to be encapsulated in an input distribution for the
integrated gluon,
$xg (x, k_0^2)$.  In Section 3 we introduce the equation for the
quark singlet
(momentum) distribution $\Sigma (x, Q^2)$, again paying
particular attention
to isolate the contribution to the non-perturbative region and to
ensure that the perturbative
terms are allowed to contribute for {\it all} $Q^2$.  In Section
4 we 
numerically solve the coupled integral equations for $f$ and
$\Sigma$.  The
$k_T$ factorization theorem is used to calculate $F_2 (x, Q^2)$
as a function of
the two parameters that specify the input gluon (which we take to
be \lq flat'
at small $x$).  Optimum fits to the available $F_2$ data at small
$x$ are
presented, and predictions are made for the charm component
$F_2^c$ and 
for the longitudinal structure function $F_L$. Finally, in
Section 5 we present our 
conclusions. \\

\noindent {\large \bf 2.  Unified BFKL and GLAP equation for the
gluon}

We start from the BFKL equation for the unintegrated gluon
distribution
\bea
\label{eq:z4}
f(x, k^2) & = & f^{(0)} (x, k^2) \nonumber \\
& & \\
& & + \: \overline{\alpha}_S (k^2) k^2 \int_x^1 \frac{dz}{z} \int
\frac{dk^{\prime 2}}{k^{\prime 2}} \left \{\frac{f(x/z, k^{\prime
2}) - 
f(x/z, k^2)}{| k^{\prime 2} - 
k^2 |}
\: + \: \frac{f(x/z, k^2)}{[4k^{\prime 4} + k^4]^{\frac{1}{2}}}
\right \} \nonumber
\eea
where $\overline{\alpha}_S = 3\alpha_S/\pi$ and $k \equiv k_T,
k^{\prime}
\equiv k_T^{\prime}$
denote the transverse momenta of the gluons, see Fig.~1. The
equation corresponds
to the leading $\ln (1/x)$ approximation. \\

\noindent {\bf 2.1~~From the BFKL to the unified equation}

In order to make the BFKL equation for the gluon 
more realistic and to extend its validity to cover the full range
of $x$ we make the following
modifications.  First, to incorporate leading order GLAP
evolution, we add on 
to the right-hand side of 
(\ref{eq:z4}) the term \cite{GLR} 
\bea
\label{eq:z5}
& & \overline{\alpha}_S (k^2) \int_x^1 \frac{dz}{z} \left
(\frac{z}{6} P_{gg} 
(z) - 1 \right ) \: \frac{x}{z} \: g \left (\frac{x}{z}, k^2
\right ) \nonumber \\
& & \\ 
& & \equiv \; \overline{\alpha}_S (k^2) \int_x^1 
\frac{dz}{z} \left(\frac{z}{6} P_{gg} (z) - 1 \right ) 
\left \{\frac{x}{z} \: g \left(\frac{x}{z}, k_0^2 \right ) \: +
\: \int_{k_0^2}^{k^2} 
\frac{dk^{\prime 2}}{k^{\prime 2}} f \left(\frac{x}{z}, k^{\prime
2} \right) 
\right \}, \nonumber 
\eea
where we have used (\ref{eq:z1}).  The --1 allows for the
contribution which is
already included in the BFKL equation. 
The inclusion of the additional term (\ref{eq:z5}) gives
contributions to the 
gluon anomalous dimension $\gamma_{gg}$ which are subleading in 
$\alpha_S \ln (1/x)$ but leading in $\alpha_S$.  These  standard
leading order 
GLAP contributions have an impact on the gluon intercept.  They
soften the small $x$ rise of the gluon distribution and also
change its overall
normalisation.

The second modification to (\ref{eq:z4}) is the introduction of
the kinematic 
constraint \cite{KMS2,AGS} 
\be
k^{\prime 2} \: < \: k^2/z
\label{eq:z7}
\ee
on the real gluon emission term, that is, on the integral over $f
(x/z, k^{\prime 2})$.  
The origin of the constraint is the requirement that the
virtuality of the
exchanged gluon is dominated by its transverse momentum,
$|k^{\prime 2} | \simeq 
k_T^{\prime 2}$.  For clarity we have restored the subscript $T$
in this equation.
The constraint is another physically motivated, subleading
correction in 
$\alpha_S \ln(1/x)$.

Thirdly, we notice that the integration region over $k^{\prime
2}$ in (\ref{eq:z4})
extends down to $k^{\prime 2} = 0$ where we expect that
non-perturbative effects will affect the
behaviour of $f (x, k^{\prime 2})$.  We are only going to solve
equation (\ref{eq:z4}) 
in the perturbative region, defined by $k^2 > k_0^2$, so we only
have to worry
about the infrared contribution due to the real emission term
from the interval
$0<k^{\prime 2}<k_0^2$.  We may rewrite this infrared
contribution in the form
\be
k^2 \int_0^{k_0^2} \frac{dk^{\prime 2}}{k^{\prime 2} | k^{\prime
2} - k^2 |}
f \left(\frac{x}{z}, k^{\prime 2} \right ) \; \simeq \;
\int_0^{k_0^2}
\frac{dk^{\prime 2}}{k^{\prime 2}} f \left(\frac{x}{z}, k^{\prime
2} \right )
\; \equiv \; \frac{x}{z} g \left(\frac{x}{z}, k_0^2 \right ).
\label{eq:z8}
\ee
The parameter $k_0^2 (\equiv Q_0^2)$ denotes the border between
the perturbative
and non-perturbative regions.  Its magnitude will be taken to be
around 
$1{\rm GeV}^2$.

Finally we must of course add to the right-hand side of
(\ref{eq:z4}) the term
which allows the quarks to contribute to the evolution of the
gluon, that is
\be
\frac{\alpha_S (k^2)}{2\pi} \int_x^1 dz P_{gq} (z) \Sigma
\left(\frac{x}{z}, k^2 \right )
\label{eq:z9}
\ee   
where $\Sigma$ is the singlet quark momentum distribution.  To be
explicit 
\bea
\label{eq:z10}
& & \Sigma (x, k^2) \; = \; \sum_{q=u,d,s} x(q + \overline{q}) \:
+ \:
x(c + \overline{c}) \nonumber \\
& & \\ 
& & \equiv V (x, k^2) \: + \: S_{uds} (x, k^2) \: + \: S_c (x,
k^2) \nonumber
\eea
where $V$, $S_{uds}$ and $S_c$ denote the valence, the light sea
quark and the
charm quark contributions respectively.  We discuss the evolution
equation for 
$\Sigma (x, k^2)$ in the next section.

Gathering together all the above modifications, equation
(\ref{eq:z4}) for 
the gluon becomes
\bea
\label{eq:z11}
& & f(x, k^2) \; = \; \tilde{f}^{(0)} (x, k^2) \nonumber \\
& & + \: \overline{\alpha}_S (k^2) k^2 \int_x^1 \frac{dz}{z} 
\int_{k_0^2} \frac{dk^{\prime 2}}{k^{\prime 2}} \left\{\frac{f
\left( {\displaystyle \frac{x}{z}}, 
k^{\prime 2} \right) \Theta \left({\displaystyle \frac{k^2}{z}} -
k^{\prime 2}\right) - 
f \left({\displaystyle \frac{x}{z}}, k^2\right)}
{| k^{\prime 2} - k^2 |} \; + \; \frac{f \left({\displaystyle
\frac{x}{z}}, k^2 \right)}{[4k^{\prime 4} 
+ k^4]^{\frac{1}{2}}} \right \} \nonumber \\
& & \nonumber \\
& & + \: \overline{\alpha}_S (k^2) \int_x^1 \frac{dz}{z}
\left(\frac{z}{6}
P_{gg} (z) - 1 \right ) \int_{k_0^2}^{k^2}  \frac{dk^{\prime 2}}
{k^{\prime 2}} f \left(\frac{x}{z}, k^{\prime 2} \right ) \: + \:
\frac{\alpha_S (k^2)}{2\pi} \int_x^1 dz P_{gq} (z) \Sigma 
\left(\frac{x}{z}, k^2 \right ), \nonumber \\
\eea
where now the driving term has the form
\be
\tilde{f}^{(0)} (x, k^2) \; = \;  f^{(0)} (x, k^2) \: + \:
\frac{\alpha_S
(k^2)}{2\pi} \int_x^1 dz P_{gg} (z) \frac{x}{z} g
\left(\frac{x}{z}, k_0^2 \right).
\label{eq:z12}
\ee
In (\ref{eq:z11}) we include the constraint $k^{\prime 2} >
k_0^2$ on the virtual,
as well as the real, contributions in order to avoid spurious
singularities at
$k^2 = k_0^2$.  In the perturbative region, $k^2>k_0^2$, we may
safely neglect the 
genuinely non-perturbative contribution $f^{(0)} (x, k^2)$ which
is expected
to decrease strongly with increasing $k^2$.  It is important to
note that we 
have avoided the necessity to parametrize $f (x, k^2)$ in the
non-perturbative 
region.  Equation (\ref{eq:z11}) only  involves $f (x, k^2)$ in
the perturbative 
domain, $k^2>k_0^2$.  The input (\ref{eq:z12}) is provided by the
conventional 
gluon at the scale $k_0^2$.  That is the input to our \lq unified
BFKL + GLAP' 
equation is determined by the same distribution as in
conventional GLAP evolution. 
The modifications to (\ref{eq:z4}) allow us to overcome the
serious limitations
discussed in the introduction.  Surprisingly, we find that we can
achieve an
excellent description of all the deep inelastic data using the
most economical
parametrization of the input gluon
$$
xg (x, k_0^2) \; = \; N(1 - x)^\beta.
$$
In particular the observed growth in $F_2 (x, Q^2)$ with
decreasing $x$ is generated
entirely by {\it perturbative} ($\ln (1/x)$ and $\ln Q^2$)
dynamics.

It is easy to see how eq.  (\ref{eq:z11})  reduces to the
conventional GLAP evolution 
equation for the gluon in the leading $\ln Q^2$ (or rather $\ln
k^2$) approximation. 
The leading $\ln k^2$ terms arise from the strongly ordered
configuration, 
$k^2_0 \ll k^{\prime 2 } \ll k^2$,     
for the real emission contributions and to the neglect of the
virtual contributions.
Then (\ref{eq:z11}) becomes
\bea
\label{eq:z16}
f (x, k^2) & = & \frac{\alpha_S (k^2)}{2\pi} \int_x^1 dz \:
P_{gg}
(z) \left[\frac{x}{z} \: g \left (\frac{x}{z}, k_0^2 \right ) \:
+ \: \int_{k_0^2}^{k^2} 
\frac{dk^{\prime 2}}{k^{\prime 2}} \: f \left (\frac{x}{z},
k^{\prime 2} \right ) \right ] 
\nonumber \\
& & \\ 
& & + \: \frac{\alpha_S (k^2)}{2\pi} \int_x^1 dz P_{gq} (z)
\Sigma
\left (\frac{x}{z}, k^2 \right ), \nonumber
\eea
where we have taken into account (\ref{eq:z12}) and the remarks
concerning the
omission of $f^{(0)}$.  Upon using (\ref{eq:z1}) we see that
(\ref{eq:z16}) 
becomes
\bea
\label{eq:z17}
\frac{\partial (xg (x, k^2))}{\partial \ln k^2} \; = \;
\frac{\alpha_S (k^2)}
{2\pi} \int_x^1 dz \left [P_{gg} (z) \frac{x}{z} \: g \left
(\frac{x}{z}, k^2 \right ) \: + \: P_{gq} (z) \Sigma \left
(\frac{x}{z}, k^2 \right ) \right ],
\eea
which is simply the GLAP evolution equation for the gluon. \\

\noindent {\bf 2.2~~Anomalous dimension of the gluon.}

We will solve (\ref{eq:z11}) for the unintegrated gluon. However
first we
anticipate the general behaviour of the anomalous dimension of
the 
gluon which will come from this equation. To do this we rewrite
the 
equation in terms of the moment function
\be
\overline{f} (\omega, k^2) \; = \; \int_0^1 dx x^{\omega-1} f (x,
k^2).
\label{eq:R1}    
\ee
We have
\bea
\label{eq:z17a}
& &\overline{f} (\omega, k^2) \; = \; \overline{f}^{(0)} (\omega,
k^2) \; + \; 
\frac{\overline{\alpha}_S (k^2)}{\omega} \: k^2
\int_{k_0^2}^{\infty}
\: \frac{dk^{\prime 2}}{k^{\prime 2}} \nonumber \\
& & \nonumber \\
& & \left \{\frac{\overline{f} (\omega, k^{\prime 2}) \left
[\Theta (k^2 - k^{\prime 2}) \:
+ \: \left (k^2/k^{\prime 2} \right )^{\omega} \Theta (k^{\prime
2} - k^2) \right ] 
- \overline{f} (\omega, k^2)}{|k^{\prime 2} - k^2|} \;
+ \; \frac{\overline{f} (\omega, k^2)}{\left (4k^{\prime 4} + k^4
\right ) ^{\frac{1}{2}}}
\right \} \nonumber \\
& &  \nonumber \\
& & + \: \overline{\alpha}_S (k^2) \: P (\omega)
\int_{k_0^2}^{k^2}
\frac{dk^{\prime 2}}{k^{\prime 2}} \overline{f} (\omega,
k^{\prime 2})                                                      
\eea
where we have neglected, for simplicity, the contribution coming
from the quarks
and where $P(\omega)$ is the moment function of 
$(zP_{gg} (z) / 6 - 1)$. 
The term in square brackets is due to the kinematic constraint.
Without this
constraint we would have 1 instead of $(k^2 / k^{\prime
2})^{\omega}$, and the two 
$\Theta$ functions would simply sum to unity.

For large $k^2$ the moment function behaves as
\be
\overline{f} (\omega, k^2) \sim \left (\frac{k^2}{k_0^2} \right
)^{\gamma_{gg} 
(\omega, \overline{\alpha}_S)}
\label{eq:z18}
\ee                                                          
where, for illustration, we take fixed $\alpha_S$. The quantity
$\gamma_{gg}$
is the anomalous dimension of the gluon. If we insert
(\ref{eq:z18}) into 
(\ref{eq:z17a}) then we find, after some algebra, the following
implicit equation 
for $\gamma_{gg}$
\be
1 - \frac{\overline{\alpha}_S}{\omega} K (\gamma_{gg}, \omega) \:
- \: 
\frac{\overline{\alpha}_S}{\gamma_{gg}} \: P (\omega) \; = \; 0 
\label{eq:z18a}
\ee

where $K$, the double moment of the kernel in (\ref{eq:z17a}), is
given by
\be
K (\gamma, \omega) \; = \; \int_0^{\infty} \frac{d \rho}{\rho}
\left \{\frac{[\rho^{\gamma} 
\Theta (1 - \rho) + \rho^{\gamma - \omega} \Theta (\rho -
1)]-1}{| \rho -1|} \;
+ \; \frac{1}{[4 \rho^2 + 1]^\frac{1}{2}} \right \}.
\label{eq:z19}
\ee
If $\omega = 0$ then the expression in square brackets reduce to
$\rho^{\gamma}$ and
we have the familiar BKFL result
\be
K(\gamma, \omega = 0) \; = \; [2 \Psi (1) - \Psi(\gamma) - \Psi
(1 - \gamma)]
\label{eq:z20}
\ee
where $\Psi$ is the logarithmic derivative of the Euler gamma
function. \\

It is clear that $\gamma_{gg}$, 
which satisfies (\ref{eq:z18a}), is of the form\footnote{If we
were to replace
$\gamma^{\rm BFKL}$ simply by the term which is leading order in
$\alpha_S$, that is
$\overline{\alpha}_S / \omega$, then the sum of the first two
terms of (\ref{eq:R7})
gives the conventional GLAP anomalous dimension.}
\be
\gamma_{gg} (\omega,\overline{\alpha}_S) \; = \;
\gamma^{\rm{BFKL}}\left(
\frac{\overline{\alpha}_S}{\omega}\right) 
\; + \; \overline{\alpha}_S \: P (\omega) \; + \; {\rm higher \:
order \: terms},
\label{eq:R7}
\ee 
where $\gamma^{\rm{BFKL}}$ satisfies the usual equation
\be
1 \: - \: \frac{\overline{\alpha}_S}{\omega} \: K
\left(\gamma^{\rm{BFKL}},\omega = 0\right) \; = \; 0.
\label{eq:R8}
\ee
The higher order terms include contributions which are subleading
in 
$\overline{\alpha}_S / \omega$ as well as in
$\overline{\alpha}_S$.
The anomalous dimension $\gamma_{gg}$ has a branch point
singularity in the 
$\omega$ plane, whose position $\omega = \omega_0
(\overline{\alpha}_S)$ controls the
small $x$ behaviour of the gluon distribution. The inverse of
(\ref{eq:R1}) gives
\be
f(x,k^2) \; \sim \; x^{-\omega_0(\overline{\alpha}_S)}.
\label{eq:R9}
\ee
The value of $\omega_0$ is obtained from the requirement that
\be
\frac{\partial}{\partial \gamma}  \left\{ 1\; - \:
\frac{\overline{\alpha}_S}
{\omega_0} \: K (\gamma,\omega_0) \: - \:
\frac{\overline{\alpha}_S}{\gamma}
\: P (\omega_0) \right\} \; = \: 0,
\label{eq:R10}
\ee
together with the equation
\be
1 - \frac{\overline{\alpha}_S}{\omega_0} \: K (\gamma, \omega_0)
\: - \:
\frac{\overline{\alpha}_S}{\gamma} \: P (\omega_0) \; = \; 0,
\label{eq:R10a}
\ee
see (\ref{eq:z18a}).
Recall that in the leading $\ln(1/x)$ (or leading $1/\omega$)
approximation (\ref{eq:R10}) reduces to
\be
\partial K(\gamma, 0)\: / \: \partial \gamma \; = \; 0,
\label{eq:R11}
\ee
which is satisfied when $\gamma = \frac{1}{2}$. Thus from
(\ref{eq:R8}) we obtain 
the well-known BFKL result that
\be
\omega_0 \quad = \quad \overline{\alpha}_S \:
K(\gamma={\textstyle \frac{1}{2}}, \omega = 0)
\quad = \quad \overline{\alpha}_S \: 4\ln2.
\label{eq:R12}
\ee
The relevant domain for solving the pair of equations
(\ref{eq:z18a}) and 
(\ref{eq:R8}) is $0 < \gamma < 1$ and $\omega > 0$. In this
region the Mellin
transform of the non-singular part of the gluon splitting
function satisfies
\be
\label{eq:R13}
P (\omega) \; < \; 0,
\ee
and moreover 
\be
K(\gamma,\omega) \; < \; K(\gamma,0).
\label{eq:R14}
\ee
Thus both the additional non-singular part of the GLAP splitting
function 
$(P_{gg}\: - \: 6/z)$ and the kinematic constraint (which takes
$K(\gamma,0)$ to $K(\gamma,\omega)$) tend to reduce the magnitude
of $\omega_0$ from the BFKL value shown in (\ref{eq:R12}). These
corrections are of course 
subleading in $\ln(1/x)$. Our numerical analysis with running
$\alpha_S$ 
reflects this softening of the $x^{-\omega_0}$ singular
behaviour. \\

\noindent {\bf 2.3~~The CCFM equation}

A more general treatment of the gluon ladder, which follows from 
the BKFL formalism is provided by the
Catani-Ciafaloni-Fiorani-Marchesini (CCFM) 
equation based on angular ordering of the gluon emission along
the chain \cite{CCFM,KMS,DOK}.
The equation embodies both the BKFL equation at small $x$ and the
conventional
GLAP evolution at large $x$. The unintegrated gluon distribution
$f$ now acquires a dependence on an additional scale (which we
may take to be $Q^2$) that 
specifies the maximal angle of gluon emission. The CCFM equation
has the form

\noindent $f(x,k^2,Q^2) \; = \; f^{(0)}(x,k^2,Q^2)$ 
\be
+ \: \overline{\alpha}_S \int^1_x \frac{dz}{z} \int
\frac{d^2q}{\pi q^2}
\Theta (Q - qz) \Delta_R(z,k^2,q^2) \frac{k^2}{(\mbox{\boldmath
$q$} + 
\mbox{\boldmath $k$})^2}
f \left(\frac{x}{z}, (\mbox{\boldmath $q$} + \mbox{\boldmath
$k$})^2, q^2 \right)
\label{eq:R20}
\ee
where the theta function $\Theta(Q - qz)$ reflects the angular
ordering 
constraint on the emitted gluon. The so-called \lq non-Sudakov'
form-factor $\Delta_R$ is 
given by 
\be
\Delta_R(z,k^2,q^2) \; = \; \exp \left[ -\overline{\alpha}_S
\int^1_z \frac{dz^\prime}{z^\prime}
\int \frac{dq^{\prime2}}{q^{\prime2}} \Theta(q^\prime-z^\prime q)
\Theta(k^2 - q^{\prime2}) \right].
\label{eq:R21}
\ee
Eq.\ (\ref{eq:R20}) contains only the singular $1/z$ term of the
$g \rightarrow gg$ splitting function
(which is screened by the virtual corrections contained in
$\Delta_R$). Its generalisation 
to include the remaining parts of this vertex (as well as the
quark
contributions) is possible.  Eq.\ (\ref{eq:R20}) has been solved
numerically in the
small $x$ domain and the solution for $f (x, k^2, Q^2)$ was
presented in
\cite{KMS}.  The CCFM equation, which is a generalisation of the 
BFKL equation, generates a steep $x^{-\omega_0}$ type of
behaviour\footnote{The
effect on $F_2$ is considered in \cite{KMS3}.} but $\omega_0$ now
acquires 
significant subleading $\ln (1/x)$ corrections which come from
the angular 
ordering constraint \cite{BOTT}.  The constraint also introduces
subleading terms
in the anomalous dimension
$$
\gamma_{gg} \; = \; \gamma_{gg}^{\rm BFKL}
\left(\frac{\overline{\alpha}_S}{\omega} \right)
\: + \: \overline{\alpha}_S \: h
\left(\frac{\overline{\alpha}_S}{\omega} \right)
\: + \: \ldots
$$                                                         
so the angular ordering constraint which gives rise to the CCFM
equation and the 
kinematic constraint (\ref{eq:z8}) lead to similar effects --
both give 
subleading $\ln (1/x)$ corrections to the \lq \lq QCD Pomeron
intercept" $\omega_0$
and to the gluon anomalous dimension $\gamma_{gg}$.  We found
that the kinematic
constraint overrides the angular ordering constraint except
possibly in the large
$x$ domain when $Q^2 < k^2$ \cite{KMS2}, see also \cite{AGS}. 
Thus in our 
formulation we neglect the angular ordering constraint
altogether. \\

\noindent{\large \bf 3.  The equation for the quark distribution}

At small $x$ the gluon drives the sea quark (momentum)
distribution $S$ via the $g \rightarrow q\overline{q}$
transition,
see Fig.~1.  We evaluate the effect using the $k_T$ factorization
theorem.  To be precise we use the $k_T$ factorization
prescription to
calculate observables (such as $F_2$) directly from the
unintegrated gluon
distribution $f (x, k_T^2)$.  For $F_2$ we interpret the result
in terms of
the sea quark distributions, implicity assuming the DIS scheme. 
The total sea
is the sum of the individual quark contributions
$$
S (x, Q^2) \; = \; \sum_q S_q (x, Q^2).
$$
At small $x$ the factorization theorem gives
\be
S_q (x, Q^2) \; = \; \int_x^1 \: \frac{dz}{z} \: \int
\: \frac{dk^2}{k^2} \: S_{\rm box}^q \: (z, k^2, Q^2) \: f \left
(
\frac{x}{z}, k^2 \right )
\label{eq:a13}
\ee
where $S_{\rm box}$ describes the quark box (and crossed
box) contribution shown in Fig.~1.  $S_{\rm box}$ implicitly
includes an integration over the transverse momentum, $\kappa$,
of the exchanged quark.  Indeed, evaluating the box contributions
we find
\bea
\label{eq:a18}
S_q (x, Q^2) & = & \frac{Q^2}{4\pi^2} \: \int \:
\frac{dk^2}{k^4} \: \int_0^1 \: d\beta \: \int \: d^2
\kappa^\prime \alpha_S \left \{ [\beta^2 + (1 - \beta)^2 ] \:
\left (\frac{\mbox{\boldmath $\kappa$}}{D_{1 q}} \: - \:
\frac{\mbox{\boldmath $\kappa$} - \mbox{\boldmath $k$}}{D_{2q}}
\right )^2 \right . \nonumber \\
& & \nonumber \\
& + & [m_q^2 \: + \: 4Q^2 \beta^2 (1 - \beta)^2 ] \:
\left (\frac{1}{D_{1 q}} \: - \: \left . \frac{1}{D_{2q}} \right
)^2 \right \} \: f \left ( \frac{x}{z}, k^2
\right ) \Theta \left(1 - \frac{x}{z} \right)
\eea
where $\kappa^\prime = \kappa - (1 - \beta) k$ and
\bea
\label{eq:q1}
D_{1q} & = & \kappa^2 \: + \: \beta (1 - \beta) Q^2 \: + \: m_q^2
\nonumber \\
& & \nonumber \\
D_{2q} & = & (\mbox{\boldmath $\kappa$} - \mbox{\boldmath
$k$})^2 \: + \: \beta (1 - \beta) Q^2 \: + \: m_q^2
\nonumber \\
& & \nonumber \\
z & = & \left [ 1 \: + \: \frac{\kappa^{\prime 2} +
m_q^2}{\beta (1 - \beta) Q^2} \: + \: \frac{k^2}{Q^2} \right ]^{-
1}. 
\eea
The argument of $\alpha_S$ is taken as $(k^2 +
\kappa^{\prime 2}) + m_q^2$.  We set the quark masses to be $m_u
= m_d = m_s = 0$ and $m_c = 1.4$ GeV. \\

\noindent{\bf 3.1~~The light quark component of the sea}

We first discuss the calculation of the contribution of the \lq
\lq
massless" $u,d,s$ quarks to the total sea distribution $S$.  It
is
necessary to consider three different regions of the $k$ and
$\kappa^{\prime}$
integrations of (\ref{eq:a18}).
\begin{itemize}
\item[(a)] The contribution from the non-perturbative region
$k^2, \kappa^{\prime 2} < k_0^2$ is evaluated phenomenologically
assuming
that it is dominated by \lq\lq soft" Pomeron exchange \cite{DL}. 
The
contribution is parametrized by the form
\be
S^{(a)} \; = \; S_u^\funp \: + \: S_d^\funp \: + \: S_s^\funp
\label{eq:a14}
\ee
where
\be
S_u^\funp \; = \; S_d^\funp \; = \; 2S_s^\funp \; = \; C_\funp \:
x^{-0.08} \: (1 - x)^8.
\label{eq:a15}
\ee
The coefficient $C_\funp$ is independent of $Q^2$ (in the large
$Q^2$ region) since the contribution arises from the region in
which the struck quarks have limited transverse momentum,
$\kappa^2 < k_0^2$.

\item[(b)] In the region $k^2 < k_0^2 < \kappa^{\prime 2}$ we
apply the
strong-ordered approximation at the quark-gluon vertex and take
\cite{BLUM}
\be
S_{\rm box} \; \rightarrow \; S_{\rm box}^{(b)} \:
(z, k^2 = 0, Q^2).
\label{eq:a16}
\ee
Then the contribution to (\ref{eq:a13}) from this domain becomes
\bea
\label{eq:a17}
S^{(b)} & = & \int_x^1 \: \frac{dz}{z} \: S_{\rm box}^{(b)} \:
(z,
k^2 = 0, Q^2) \: \int_0^{k_0^2} \: \frac{dk^2}{k^2} \:
f \left ( \frac{x}{z}, k^2 \right
) \nonumber \\
& & \\
& = & \int_x^1 \: \frac{dz}{z} \: S_{\rm box}^{(b)}
\: (z, k^2 = 0, Q^2) \: \frac{x}{z} \: g
\left ( \frac{x}{z}, k_0^2 \right ) \nonumber
\eea
where the summation over $u,d,s$ is implicitly assumed.
The potential collinear singularities in the on-shell
structure function $S_{\rm box}$ are regulated by the
cut-off $k_0^2$.  Recall that $\kappa^2 \simeq \kappa^{\prime 2}
>
k_0^2$.

\item[(c)] In the remaining region, $k^2 > k_0^2$, eq.\
(\ref{eq:a18}) is left unchanged.  To be precise we use the
perturbative expression for $S_q (x, Q^2)$. 
\end{itemize}

\noindent{\bf 3.2~~The charm component}

The calculation of the charm component of the sea follows
perturbative
QCD in all regions.  To evaluate $S_{q=c}$ we divide the
integration over
$k^2$ into the regions $k^2 < k_0^2$ and $k^2 > k_0^2$.  For $k^2
< k_0^2$, 
which we denote region (b), we use the on-shell approximation to
evaluate
$S_{\rm box}$.  That is we calculate $S_{\rm box} (z, k^2 = 0,
Q^2; m_c^2)$, which
is finite due to $m_c \neq 0$.  Then (\ref{eq:a13}) gives
\be
S_{q=c}^{(b)} (x, Q^2) \; = \; \int_{x}^a \frac{dz}{z} \: S_{\rm
box} 
(z, k^2 = 0, Q^2; m_c^2) \int_0^{k_0^2} \frac{dk^2}{k^2} \: f
\left(\frac{x}{z}, 
k^2 \right)
\label{eq:t2}
\ee
where $a \; = \; (1 + 4m_c^2 / Q^2)^{-1}$, see (\ref{eq:q1}). 
For $k^2 > k_0^2$, which we call region
(c), we use the full perturbative formula.  Thus adding the two
contributions
we have
\bea
\label{eq:t3}
S_{q=c} (x, Q^2) & = & \int_{x}^a \frac{dz}{z} \: S_{\rm box} (z,
k^2 = 0,
Q^2; m_c^2) \frac{x}{z} \: g \left(\frac{x}{z}, k_0^2 \right)
\nonumber \\
& & + \: \int_{x}^a \frac{dz}{z} \: \int_{k_0^2} \frac{dk^2}{k^2}
\:
S_{\rm box} (z, k^2, Q^2; m_c^2) f \left(\frac{x}{z}, k^2
\right),
\eea
where we have used (\ref{eq:z1}) which enables $S_{q=c}$ is to be
specified
in terms of the conventional gluon input distribution. \\

\noindent{\bf 3.3~~The equation for the quark singlet
distribution}

Besides $S$, the singlet momentum distribution $\Sigma$ also
contains a valence quark contribution $V$, which is taken from a
known set of partons.  Thus in summary the singlet distribution
is
\be
\Sigma \; = \; (S^{(a)} \: + \: S^{(b)} \: + \: S^{(c)})_{uds}
\: + \: (S^{(b)} + S^{(c)})_{q=c} \: + \: V,
\label{eq:a19}
\ee
where $S^{(a)}$ is phenomenologically parametrized in terms of
\lq\lq soft" Pomeron exchange and the $S^{(b)}$ terms are
determined
perturbatively except for the (non-perturbative) input gluon
distribution 
at scale $k_0^2$. The $S^{(c)}$ terms are defined entirely in
terms of the 
unintegrated gluon distribution $f$ in the perturbative region
$k^2 > k_0^2$. 
Finally, $V = x (u_{\rm val} + d_{\rm val})$ is the valence quark
contribution.

In order to see the connection with the GLAP evolution of the
(light) quark sea we first note that
\be
S_q (x, Q^2) \; = \; S_q (x, k_0^2) \: + \: \int_{k_0^2}^{Q^2}
\: \frac{\partial S_q (x, Q^{\prime 2})}{\partial Q^{\prime 2}}
dQ^{\prime 2}, 
\label{eq:a20}
\ee
where here $S$ denotes the sum over just the $u$, $d$ and $s$
quarks.  We next 
recall that the leading twist part of the $k_T$ factorization
formula (\ref{eq:a13}), 
written in the form
\be
Q^2  \frac{\partial S_q (x, Q^2)}{\partial Q^2} \; =
\; \int_x^1 \: \frac{dz}{z} \: \int \: \frac{dk^2}{k^2} \: Q^2
\frac{\partial S_{\rm box}^q (z, k^2, Q^2)}{\partial
Q^2} \: f \left ( \frac{x}{z}, k^2 \right ),
\label{eq:a21}
\ee
can be reduced to the collinear form \cite{KM}
\be
Q^2 \frac{\partial S_q (x, Q^2)}{\partial Q^2} \; =
\; \frac{\alpha_S (Q^2)}{2\pi} \: \int_x^1 \:
dz \: P_{qg} (z, \alpha_S
(Q^2)) \: \frac{x}{z} \: g \left ( \frac{x}{z}, Q^2
\right )
\label{eq:a22}
\ee
which incorporates leading $\ln 1/x$ resummation effects in both
the splitting function $P_{qg}$ and in the integrated gluon
distribution $g$.  Thus (\ref{eq:a20}) may be written in the form
\bea
\label{eq:a23}
S_q (x, Q^2) & = & S_q (x, k_0^2) \: + \:
\int_{k_0^2}^{Q^2} \: \frac{dQ^{\prime 2}}{Q^{\prime 2}} \:
\frac{\alpha_S (Q^{\prime 2})}{2\pi} \: \int_x^1 \:
dz \nonumber \\
& & \\
& & \left [P_{qg} (z, \alpha_S (Q^{\prime 2})) \:
\frac{x}{z} \: g \left ( \frac{x}{z}, Q^{\prime 2}
\right ) \: + \: P_{qq} (z, \alpha_S (Q^{\prime 2})) S_q
\left(\frac{x}
{z}, Q^{\prime 2} \right )  
\right ], \nonumber
\eea
where for consistency we have included the $S \rightarrow S$
contribution
to the evolution.  This additional term is needed to ensure the
correct GLAP 
structure.  Of course, at small $x$ we expect $S$ to be
dominantly driven
by the gluon.  Equation (\ref{eq:a23}) is simply the integral
form of the 
GLAP evolution equation for the (light) sea quark (momentum)
distribution, $S$.

Guided by the GLAP structure, it is clear that we should also add
the
$S \rightarrow S$ contribution to the complete equation
(\ref{eq:a19}) based
on $k_T$ factorization. Then (\ref{eq:a19}) becomes
\bea
\label{eq:z13}
\Sigma (x, k^2) & = & S^{(a)} (x) \: + \: \sum_q \int_{x}^a
\frac{dz}{z} \:
S_{\rm box}^q (z, k^{\prime 2} = 0, k^2; m_q^2) \frac{x}{z} \: g
\left(\frac{x}{z},
k_0^2 \right)
\: + \: V (x, k^2) 
\nonumber \\
& & \nonumber \\
& & + \: \sum_q \int_{k_0^2}^\infty \frac{dk^{\prime
2}}{k^{\prime 2}} 
\int_x^1 \frac{dz}{z} \: S_{\rm box}^{q} (z, k^{\prime 2},
k^2; m_q^2) f \left (\frac{x}{z}, k^{\prime 2} \right) \nonumber
\\
& & \\
& & + \: \int_{k_0^2}^{k^2} \frac{dk^{\prime 2}}{k^{\prime 2}} \:
\frac{\alpha_S
(k^{\prime 2})}{2\pi} \int_x^1 dz \: P_{qq} (z)
S_{uds} \left(\frac{x}{z}, k^{\prime 2} \right ) \nonumber
\eea
where $S^{(a)}$ is given by (\ref{eq:a14}) and
where the $uds$ subscript indicates that the additional $S
\rightarrow S$ term is
only included for the light quarks.  This equation for the
singlet quark 
distribution $\Sigma$, together with eq.\ (\ref{eq:z11}) for the
gluon, form the 
pair of coupled equations which we solve.  In this way we can
specify the structure
function $F_2$ in terms of the parameters of the input
distributions, and hence
determine the values of the parameters by fitting to the data for
$F_2$, see
sections 4-6. \\

\noindent{\bf 3.4~~$k^T$ versus collinear factorization and
$P_{qg}$}

As we have already mentioned, the {\it leading-twist} part of
the $k_T$ factorization formula can be rewritten in a collinear
factorization form. Once the unintegrated gluon distribution is
taken as a solution of the BFKL equation and the $k_T$
factorization integral is performed over the {\it entire}
available phase-space (i.e. not only over the region
corresponding to the strongly ordered transverse momenta) then 
the leading small $x$ effects are automatically resummed in
the splitting functions and in the coefficient functions.  The
$k_T$ factorization theorem can in fact be used as the tool for
calculating these quantities \cite{KTFAC,CIAFKT}.  

We illustrate this point by using the 
example of the calculation of the splitting function $P_{qg}$. 
For simplicity
we assume that the coupling $\alpha_S$ is fixed and that the
quarks are massless.
We begin from the $k_T$ factorization formula (\ref{eq:a21})
written in moment
space
\be
Q^2 \frac{\partial \overline{S} (\omega, Q^2)}{\partial Q^2}\; =
\; \int
\frac{dk^2}{k^2} \: Q^2 \frac{\partial \overline{S}_{\rm box}
(\omega, Q^2 / k^2)}
{\partial Q^2} \: \overline{f} (\omega, k^2).
\label{eq:t4}
\ee
where we have noted, for massless quarks, that $\overline{S}_{\rm
box}$ is a
function of the ratio $Q^2 / k^2$.  Thus we may factorise the
convolution over
$k^2$ by taking moments.  We find  
\begin{equation} 
Q^2{\partial \bar S(\omega, Q^2) \over  \partial Q^2} = {1\over 2
\pi i} \int_{1/2-i\infty}^{1/2+i\infty} d\gamma \gamma \tilde
S_{\rm box}(\omega,\gamma) \tilde f(\omega,\gamma)\left(Q^2\over
k_0^2 \right)^{\gamma}
\label{dconv}
\end{equation}
where  $\tilde S_q^{box}(\omega,\gamma)$ and $\tilde
f(\omega,\gamma)$ are the Mellin transform of the moment
functions $\bar S_q^{box}(\omega, k^2,Q^2)$ and $\bar
f(\omega,k^2)$ i.e. 
\begin{equation}
\bar S_{\rm box}(\omega, k^2,Q^2) \: = \: {1\over 2 \pi i}
\int_{1/2-i\infty}^{1/2+i\infty} d\gamma \tilde
S_{\rm box}(\omega,\gamma)\left(Q^2\over k^2 \right)^{\gamma}
\label{mbox}
\end{equation}
and
\begin{equation}
\bar f(\omega, k^2)={1\over 2 \pi i}
\int_{1/2-i\infty}^{1/2+i\infty} d\gamma \tilde f
(\omega,\gamma)\left(k^2\over k_0^2 \right)^{\gamma}.
\label{fmell}
\end{equation} 
Retaining the leading pole $\gamma_{gg}$ 
contribution of $\tilde f(\omega,\gamma)$ in the $\gamma$ plane, 
the integrals (\ref{dconv}) and (\ref{fmell}) can be evaluated to
give 
\begin{equation}
Q^2{\partial \bar S(\omega, Q^2) \over  \partial Q^2} \; = \;
\gamma_{gg}^{2}\tilde S_{\rm box} \left(\omega,
\gamma_{gg}\right) C(\omega,\alpha_S)\left({Q^2 \over k_0^2}
\right)^{\gamma_{gg}}
\label{eq:asympts}
\end{equation}
and 
\begin{equation}
\bar f(\omega, k^2)= \gamma _{gg} C (\omega,\alpha_S) 
\left({k^2 \over k_0^2} \right)^{\gamma_{gg}}.
\label{eq:asymptf}
\end{equation} 
where $\gamma_{gg} C$ is the residue of the pole. 
The function $\gamma_{gg}(\bar \alpha_S, \omega)$ is the (leading
twist) gluon anomalous dimension.  From (\ref{eq:asymptf}) we see
that the
integrated gluon is given by
\begin{equation} 
\bar g(\omega, Q^2)=C (\omega,\alpha_S) \left({Q^2 \over k_0^2}
\right)^{\gamma_{gg}} 
\label{grep}
\end{equation}
where $\overline{g}(\omega,Q^2)$ is the moment function of the
(leading twist
part) of the gluon distribution.
Thus by comparing (\ref{eq:asympts})
with the conventional GLAP form 
\begin{equation}
Q^2{\partial \bar S_q(\omega, Q^2) \over  \partial Q^2} =
{\alpha_S\over 2\pi}\bar P_{qg}(\omega,\alpha_S)\bar
g(\omega,Q^2)
\label{apqg}
\end{equation}
we can identify the moment of the $P_{qg}$ splitting function to
be
\begin{equation}
{\alpha_S \over 2\pi} \bar P_{qg}(\omega,\alpha_S) = 
\gamma_{gg}^{2}(\bar \alpha_S, \omega)\tilde S_{\rm box}^q
\left(\omega, \gamma_{gg}(\bar \alpha_S, \omega) \right)
\label{eq:pqgf}
\end{equation} 
in the so-called $Q_0^2$ regularization and DIS
scheme \cite{CIAFKT} which we implicitly adopt.  
 
In the leading $\ln(1/x)$ approximation we have 
\begin{equation}
\frac{\alpha_S}{2\pi} \bar P_{qg}(\omega,\alpha_S)= 
(\gamma^{\rm BFKL})^2 \tilde S_{\rm box}^q
\left(\omega=0, \gamma^{\rm BFKL} \right).
\label{eq:pqgflx}
\end{equation} 
The anomalous dimension $\gamma^{\rm BFKL}$ has the following
expansion \cite{JAR} 
\begin{equation}
\gamma^{\rm BFKL} \left(\frac{\overline{\alpha}_S}{\omega}
\right) \; = \; 
\sum_{n=1}^{\infty}c_n
\left({\bar \alpha_S\over \omega} \right)^n
\label{eq:adexp}
\end{equation}
which in turn gives for the splitting function $P_{gg}$
\begin{equation}
zP_{gg}(z,\alpha_S)=\sum_{n=1}^{\infty}c_n{[\overline{\alpha}_S
\ln(1/z)]^{n-1}\over (n-1)!},
\label{eq:pggzet}
\end{equation}
whereas representation (\ref{eq:pqgflx}) generates the following 

expansion of the splitting function $P_{qg}(z,\alpha_S)$ at small
$z$  
\begin{equation}
zP_{qg}(z,\alpha_S) \: = \: zP_{qg}^{(0)}(z) +
\bar \alpha_s \sum_{n=1}^{\infty}b_n
{[\bar \alpha_S \ln(1/z)]^{n-1}\over (n-1)!}.
\label{eq:zpqgf}
\end{equation}
The first term on the right hand side 
vanishes at $z=0$.  It should be noted that the splitting
function $P_{qg}$ is formally non-leading at small $z$ when
compared with the splitting function $P_{gg}$.  For moderately
small values of $z$ however, when the first few terms in the
expansions (\ref{eq:zpqgf}) and (\ref{eq:pggzet}) dominate, the
BFKL
effects can be much more important in $P_{qg}$  than in $P_{gg}$.

This comes from the fact that  
all coefficients $b_n$ in (\ref{eq:zpqgf}) are different from
zero, while in
(\ref{eq:pggzet}) we have $c_2=c_3=0$ \cite{JAR}.  The small $x$
resummation effects within the conventional QCD evolution 
formalism have recently been discussed in refs.\
\cite{EKL,HBRW,BFORTE,FRT,BRV}.  These studies already
emphasize this point, namely
that at the moderately small values
of $x$ which are relevant for the HERA measurements, the 
$\ln (1/x)$ resummation effects in the splitting function
$P_{qg}$ have a
much stronger impact on $F_{2}$ than do those in
the splitting function $P_{gg}$.  In particular we should also
recall
that the BFKL effects in the splitting function
$P_{qg}$ can significantly affect the extraction of the
gluon distribution from the experimental data on the slope of
the structure function $F_2$ 
\begin{equation}
Q^2{\partial F_2(x,Q^2)\over \partial Q^2} \simeq 
\sum_q e_q^2 \frac{\alpha_S(Q^2)}{2\pi} \:
\int_x^1 dz P_{qg}(z,\alpha_S(Q^2)){x\over z}
g \left({x\over z},Q^2 \right). 
\label{slopef2}
\end{equation}
Here we also include the subleading $\ln(1/x)$ terms
which would come from the subleading terms in $\gamma_{gg}$ etc. 
Keeping the exact $k_T$ factorisation (and not just its large
$Q^2$ limit) we also include the non-leading twist
contributions to $F_2$.  They would formally be generated
by the contributions given by the (non-leading) twist anomalous
dimensions. \\

\noindent {\large \bf 4.  Numerical analysis and the description
of $F_2$}

We now have a closed system of two coupled integral equations for
two unknowns.  
Namely equation (\ref{eq:z11}) of Section 2 for the unintegrated
gluon distribution
$f (x, k^2)$ and equation (\ref{eq:z13}) of Section 3 for the
integrated quark
singlet (momentum) distribution $\Sigma (x, k^2)$.  The effect of
the gluon in the
perturbative region, $k^2>k_0^2$, is of special interest.  It is
the \lq dynamo'
which drives small $x$ physics.

The advantage of this formulation of the unified BFKL/GLAP
equation is that the
input is well-controlled.  We emphasized in Section 2 that the
equation for
$f (x, k^2)$ required only the specification of an input form for
the integrated
gluon,
\be
xg (x, k_0^2) \; = \; N (1-x)^\beta,
\label{eq:z14}
\ee
say. Moreover, the equation for the singlet $\Sigma (x, k^2)$
requires as input
only the non-perturbative sea contribution whose form we assume
is given by
the \lq \lq soft" Pomeron
\be
S^{(a)} \; = \; C_\funp x^{-0.08} (1-x)^8
\label{eq:z15}
\ee
and the contributions $S^{(b)}$ of (\ref{eq:a17}) and
(\ref{eq:t2})
which depend on $xg (x, k_0^2)$ of (\ref{eq:z14}).  The choice of
the exponent 
--0.08 is motivated by the Regge pomeron intercept found in the
analysis of total
cross section data \cite{DL}.  We choose the exponent of $(1 -
x)$ to be 8, 
typical of the behaviour of the sea distribution.  In our small
$x$ analysis any
similar choice would be equally good and would not change the
quality of the
description.

The valence quark contribution $V(x, k^2)$ in (\ref{eq:z13}),
which is determined
mainly by fixed target deep inelastic data, is taken from the
leading order 
GRV set of partons \cite{GRV}.  We are therefore able to
self-consistently 
determine $f (x, k^2)$ and $\Sigma (x, k^2)$ as functions of a
small
number of physically motivated parameters.  In fact, we have only
the two 
parameters, namely $N$ and $\beta$ 
determining the input gluon distribution (\ref{eq:z14}). The
momentum sum 
rule fixes the value of $C_\funp$, which determines the input
sea, 
(\ref{eq:z15}).  The presence of BFKL - like terms means that the
momentum sum
rule is not exactly conserved.  However the violation is small. 
For example, after evolution to
$Q^2 \: = \: 50 {\rm GeV^2}$ we find that the sum of the momentum 
fractions carried by the gluon and the light quarks
is only increased from 1 to 1.007. We neglect this small
violation of momentum conservation. \\

\noindent {\bf 4.1~~The optimum description of the $F_2$ data at
small $x$}  

We determine the values of the input parameters by fitting to the
HERA
measurements of the proton structure function $F_2$ using
\be
\label{eq:a25}
F_2 \; = \; \sum_q \: e_q^2 (S_q + V_q), 
\ee
which holds in the DIS scheme. We thus have to calculate $S_q (x,
Q^2)$ in terms of the
input gluon parameters $N$ and $\beta$.  To do this we
solve the pair of equations (\ref{eq:z11}) and (\ref{eq:z13}) for
$f (x, k^2)$ and $\Sigma (x, k^2)$ using an extension of
the method proposed in \cite{czebmet}. This method incorporates
the interpolation in two variables $x$ and $Q^2$ with orthogonal
polynomials. Thus the coupled integral equations 
can be transformed into the set of linear algebraic equations and
readily
solved.  In this way we can express $F_2 (x,Q^2)$ in terms of $N$
and $\beta$.  
We then determine the optimum values of these parameters by
fitting to the HERA
\cite{HERA} and fixed-target \cite{ENB} data for $F_2 (x, Q^2)$
that are available in 
the small $x$ domain, $x < 0.1$.  We take a running coupling
which satisfies 
$\alpha_S (M_Z^2) \: = \: 0.12$.

We actually show the results of two fits. The first is the \lq
realistic' fit
with the kinematic constraint imposed (which requires the
virtuality of the
exchanged gluons along the ladder to satisfy $|k^{\prime 2}|
\simeq k_T^{\prime 2})$.
Then for comparison we repeat the analysis without imposing the
kinematic constraint, 
that is we omit $\Theta$ function in (\ref{eq:z11}).  The quality
of the fits
are shown in
Figs.\ 2 and 3, and the parameters given in Table 1. To be
precise Figs.~2 and 3
respectively show the description of the H1 and ZEUS data
\cite{HERA}
together with those fixed-target data that occur at the same
values of $Q^2$.

The fit with the kinematic constraint included (continuous
curves) is significantly
better than that in which it is omitted (shown by the dashed
curves).  Without
the constraint the predicted rise of $F_2$ is a little too steep
at the smallest
values of $x$ and $Q^2$.  Over the remainder of the $x, Q^2$
domain the fit
(fit~2) gives a good description of $F_2$.  It is far better, for
example, than
that shown in ref. \cite{BE}.

The kinematic constraint, which corresponds to subleading
$\ln (1/x)$ corrections, lowers the \lq hard' pomeron intercept
and improves the
description of the data, particularly at the smaller values of
$x$.  In fact the
resulting description of $F_2 (x, Q^2)$
with just two free parameters ($N$ and $\beta$) is excellent, and
is comparable, even a little better than, 
to that achieved in the global parton analyses, see, for example,
the $\chi^2$
listed in Table 1.  Moreover, the overall behaviour of the gluon
is much more 
realistic than that of
the fit without the kinematic constraint.  It gives an acceptable
description of 
the WA70 prompt photon data \cite{WA70}, which directly sample
the gluon at
$x \simeq 0.4$.
These data were not used to constrain the gluon. For fit 1 the
prediction is some $30\%$ above WA70 data, which is within
the QCD scale uncertainties, whereas the gluon of fit 2
gives a prediction which is a factor of about 2.5 above the data.
It is not surprising that the gluons are so different in the two
fits since they are both contrived to give satisfatory description
of the measurements of $F_2$ at small $x$, despite the fact that the
kinematic constraint significantly reduces the gluon intercept
$\omega_0$. It is encouraging that it is the description with the
kinematic constraint that gives the acceptable large $x$ behaviour
of the gluon.  
For completeness we use our determination of the unintegrated
gluon to compute
the conventional gluon distribution $xg (x, Q^2)$ and compare the
result with the
gluons of recent sets of partons obtained in GLAP-based global
analyses of deep
inelastic and related data.  To be specific the continuous curves
in Fig.~4 compare the 
gluon calculated from $f (x, k_T^2)$ of the fit~1 (via eq.
(\ref{eq:z1})) with
the gluon distributions of the MRS(R2) \cite{MRS} and GRV
\cite{GRV} set of partons
(shown by the dashed and dotted curves respectively).  We see
that the behaviour
of the integrated gluon is very similar to that of MRS(R2).  This
may be expected
since the MRS analysis used to the same HERA data as those fitted
in the 
present work, whereas these data were not available at the time
of the GRV analysis.  However,
we emphasize the different underlying structure of the present
analysis and the
pure GLAP-based descriptions.  We shall see below that in the
unified BFKL/GLAP
approach the rise of $F_2$ is generated essentially by $\ln
(1/x)$ effects in
the off-shell gluon structure, $F_2^{\gamma g}$ of Fig.~1.

\begin{table} [htb]
\begin{center}
\begin{tabular}{|l|c|cc|c|c|} \hline
& Kinematic & \multicolumn{2}{|c|}{$xg = N(1 - x)^{\beta}$}& &
$\chi^2/{\rm data point}$ \\
& constant & $N$ & $\beta$ & $C _{\funp}$ & [393 points] \\
\hline
Fit 1 & yes & 1.57 & 2.5 & 0.269 & 1.07 \\
Fit 2 & no & 0.85 & 0.9 & 0.269 & 1.8 \\ \hline
MRS(R2) & & & & & 1.12 \\ \hline
\end{tabular}
\end{center}
\end{table}
\noindent Table 1:  The parameters $N$ and $\beta$ determined in
the optimum fit to the
available data \cite{HERA,ENB} for $F_2$ with $x < 0.05$ and $Q^2
> 1.5 {\rm GeV^2}$, 
without and with the inclusion of the kinematic constraint
along the gluon ladder.  The value of $C _{\funp}$ of
(\ref{eq:a15}) is also
shown, although this is fixed in
terms of $N$ and $\beta$ by the momentum sum rule.  For
comparison we also show
the $\chi^2$ for the same set of HERA and fixed-target data
obtained in a recent next-to-leading
order GLAP global parton analysis \cite{MRS}. 
 For both fit 1 and 2 the gluon carries $45\%$ of the proton's
 momentum at the input scale $k_0^2=1 GeV^2$. \\

\noindent{\bf 4.2~~The effect of the $\ln (1/x)$ resummation on
the gluon}

Fig.~5 shows the behaviour of the unintegrated gluon distribution
$f (x, k^2)$
as a function of $k^2$ for $x = 10^{-3}$ and $10^{-4}$.  Three
different 
determinations are shown, each of which start from the input
$$
xg (x, k_0^2) \: = \: 1.57 (1-x)^{2.5}
$$
of fit~1 of Table~1.  The continuous and dashed curves
correspond, respectively, to
the behaviour with and without the kinematic constraint.  The
dotted curve is
obtained from a GLAP determination in which the BFKL kernel in
(\ref{eq:z11}) is
replaced by the leading order $P_{gg}$ function.  That is
(\ref{eq:z11}) is
replaced by
\begin{eqnarray}
\label{eq:q2}
f(x,k^2) & = & {{\alpha}_S(k^2) \over 2\pi}\left[
\int_x^1 {dz} {P}_{gg}(z)
{x \over z} g \left({x \over z}, k_0^2 \right) \right. \nonumber
\\ 
& & + \: \left.\int_{x}^{1}{dz} {P}_{gg}(z) \int_{k_0^2}^{k^2}
{dk'^2
\over
k'^2} f \left({x \over z},k'^2 \right) + \int_{x}^1 { dz }
{P}_{gq}(z)
\Sigma \left({x \over z},k^2 \right) \right],
\end{eqnarray}
where $P_{gg}$ has the usual form
\begin{eqnarray}
P_{gg}(z) = 6\left[{1-z \over z} + z (1-z) + {z \over (1-z)_{+} }
+
{11 \over 12}
\delta(1-z)\right]-{N_f \over 3}\delta(1-z). 
\end{eqnarray}
The comparison of the dashed and dotted curves shows
that the differences between the BFKL and
GLAP approaches are not very big,  even for the values of $x$ as
low
as $10^{-3}$.
The differences become more obvious when one considers smaller
values of $x$,
around $10^{-4}$. Even then the discrepancies are only visible at
lower
values of $k^2$.
This effect can be explained in terms of power series expansion
in
$\alpha_S / \omega$
of the gluon anomalous dimension
$$
\gamma ^{\rm BFKL} \: = \: \frac{\overline{\alpha}_S}{\omega} \:
+ \:
c_4 \left(\frac{\overline{\alpha}_S}{\omega} \right)^4 \: + \:
\ldots
$$
see (\ref{eq:adexp}), where $c_2 = c_3 = c_5 = 0$.  The first
term of the expansion,
which is common to BFKL and GLAP, is clearly dominant for the
smaller values of
$\alpha_S (k^2)$.  Thus we confirm the well known result that, in
the region of 
moderately small values of $x$ relevant for the HERA
measurements, $\ln (1/x)$
resummation has little effect on the gluon distribution.  If the
gluon input were
adjusted to correspond to the optimum fit with the kinematic
constraint imposed, 
then the continuous curve would be comparable to the other two. 
However, a common
input is used to show the impact of the kinematic constraint.  We
see that the
resulting gluon is smaller and less steep.  This implies that
subleading 
$\ln (1/x)$ corrections are significant.\\

\noindent {\bf 4.3~~Effect of $\ln (1/x)$ resummation on the  
structure function $F_2$}

To investigate the various effects of the $\ln (1/x)$ terms we
compute
$F_2 (x, Q^2)$ using four different procedures but with a common 
input\footnote{In this calculation we have only included the
light quarks
$u,d,s$ (which we treat as massless), since we want to avoid any
dependence
on the choice of scale for the heavy quarks. Such a dependence
would spoil the
clarity of the explanation of some effects which we want to
stress.  Everywhere 
else we include also the charm quark.}
$$
xg (x, k_0^2) \: = \: 1.57 (1 - x)^{2.5},
$$
correspoding to fit 1.
The four different determinations are shown in Fig.~6 and
correspond to
\begin{itemize}
\item[(i)] The full unified BFKL + GLAP calculation with the
kinematic constraint
included, eqs.(\ref{eq:z11}) and (\ref{eq:z13}), shown as a
continuous curve.
\item[(ii)] Analogous to (i) but without the kinematic constraint
(dashed
curve).
\item[(iii)] Replace (\ref{eq:z11}) by the pure GLAP equation in
the gluon 
sector, eq. (\ref{eq:q2}), but keep the full $k_T$ factorization
for the quarks
(dotted curve).
\item[(iv)] Pure GLAP evolution for both the gluons and the
quarks (dot-dashed
curve).  That is instead of (\ref{eq:z13}) we use
\begin{eqnarray}
\Sigma(x,k^2)=S^{(a)}(x)+S^{(b)}(x,k^2)+V(x,k^2)+\nonumber \\
\int_x^1 dz {P}_{qg}(z)
\int_{k_0^2}^{k^2}{d{k'}^2 \over {k'}^2}
f \left({ x \over z},{k'}^2 \right) \xi({k'}^2,k^2) 
\label{eq:dglapquark}
\end{eqnarray}
where $\xi({k'}^2,k^2)$ is the evolution length and is defined
by,
\begin{equation}
\xi({k'}^2,k^2)=\int_{k'^2}^{k^2} {d{q'}^2 \over {q'}^2}
{\alpha}_S({q'}^2).
\label{evlenght}
\end{equation}
\end{itemize}
 One can again see from Fig.~6 that the differences between BFKL
 (with no kinematical constraint) and GLAP evolution in the gluon
sector
 are not very big. The calculations start to differ only
 at $x\simeq 10^{-4}$ ( dashed and dotted lines).
 On the other hand  when we compare the pure GLAP evolution
 (with the $P_{qg}$ splitting function) with the equations where
 the entire phase space has been taken into account
  then the differences are much 
 bigger. This implies that the leading order terms in $\alpha_S
\ln 1/x$ present
 in the gluon off-shell structure function ($F^{\gamma g}_2$)  
 are much more important than the terms in the
gluon anomalous dimension resulting from the BFKL equation.
The effect of the kinematic constraint is again evident.  It
leads to the change
from the dashed to the continuous curves.  Fig.~6 also enables us
to see the
$x$ values at which the effect of the $\ln (1/x)$ summation
effects become
important. \\

\noindent {\bf 4.4~~Predictions for $F_2^c$ and $F_L$}

Once we have determined the parton distributions we can predict
the values of
other hard scattering observables.  At small $x$ we see, via the
$k_T$ 
factorization theorem, that the observables are \lq driven' by
the unintegrated
gluon distribution $f (x, k^2)$. Here we calculate $F_2^c$ and
$F_L$.

The charm component $F_2^c$ of $F_2$ is given by 
$$
F_2^c (x, Q^2) \: = \: e_c^2 \: S_{q=c} (x, Q^2)
$$
where the charm sea $S_{q=c}$ is calculated from (\ref{eq:t3}) in
terms of the 
gluon.  It is the second term on the right-hand side of
(\ref{eq:t3}) which
drives the small $x$ behaviour.  The predictions are compared
with the H1
measurements \cite{H1} of $F_2^c$ in Fig.~7.  The percentage of
charm in the
deep inelastic structure function is shown in Fig.~8. At small
$x$ we see that
$F_2^c$ is an appreciable fraction of $F_2$.  Recall that in the
massless charm
limit the fraction would be 0.4, provided that we are below the
bottom quark 
threshold.

The predictions of the longitudinal structure function $F_L$ are
shown in
Fig.~9.  For $F_L$ the $k_T$ factorization formula can be written
in the form
\cite{BLUM,BBJKAS} 
\begin{eqnarray}
F_L(x,Q^2) & = &
\frac{\alpha_s(Q^2)}{\pi}\left[\frac{4}{3}\int_x^1\frac{dy}{y}
\left(\frac{x}{y} \right)^2F_2(y,Q^2) \right.
 +\left.\sum_qe_q^2\int_x^1\frac{dy}{y} \left(\frac{x}{y}
\right)^2
\left(1-\frac{x}{y} \right)yg(y,k_0^2)\right]\ \nonumber \\ 
& & + \: \sum_q
e_q^2\frac{Q^4}{\pi^2}\int_{k_0^2}\frac{dk^2}{k^4}
\int_0^1d\beta\beta^2(1-\beta)^2 \:
 \int d^2 \kappa'\alpha_{S}
\left[\frac{1}{D_{1q}}-\frac{1}{D_{2q}}\right]^2 \: f \left({x
\over
z},k^2 \right),
\end{eqnarray}
where the quark box variables $D_{iq}$ and $\kappa^{\prime}$ are
defined in
(\ref{eq:q1}).  The behaviour of $F_L$ is driven by the gluon
through the last 
term.  The argument $\alpha_S$ is taken to be $k^2 +
\kappa^{\prime 2} + m_q^2$
as before. \\

\noindent {\large \bf 5.~~Conclusions}

The natural framework for describing \lq hard' scattering
observables at small
$x$ Bjorken $x$ is provided by the gluon distribution $f (x,
k_T^2)$
{\it unintegrated} over its transverse momentum, together with
the $k_T$
factorization theorem.  At small $x$ it is only to be expected
that the
$k_T$ dependence should be treated explicitly.

In the leading $\ln (1/x)$ limit, $f (x, k_T^2)$ satisfies the
BFKL equation.  To
make a smooth transition to the larger $x$ domain we have studied
a modified
equation which treats the leading $\ln (1/x)$ and the leading
$\ln (Q^2)$ terms
on an equal footing.  Moreover, we arrange the equation so that
we need only to 
consider $f (x, k_T^2)$ in the perturbative domain, $k_T^2 >
k_0^2$.  The 
integrated gluon distribution $xg (x, k_0^2)$ is the only
non-perturbative
input that is required.  At small $x$, the singlet quark
distribution 
$\Sigma (x, Q^2)$ is controlled by $f (x, k_T^2)$ through the $g
\rightarrow
q \overline{q}$ splitting.  We therefore extend the formalism to
a pair of 
coupled integral equations for $f$ and $\Sigma$ which embrace
both the BFKL
leading $\ln (1/x)$ and GLAP leading $\ln (Q^2)$ contributions in
a consistent
way.  A notable feature of the formalism is that we can retain
the full
perturbative contribution of the quark box which contributes to
the sea
distribution for {\it all} $Q^2$.  In this way we can isolate the
non-perturbative
contribution to a (scaling) sea contribution whose general form
is known from
\lq soft' physics.

An alternative way to unify the BFKL and GLAP formalisms is based
on collinear
factorization.  It has been shown that we can reduce the (leading
twist part
of the) BFKL behaviour to collinear form in which the the
splitting and
coefficient functions contain explicit calculable series of 
$\alpha_S^m (\alpha_S \ln (1/x))^n$ terms.  This approach has
attracted much
interest. However, in the introduction we stressed the advantage
of working with the 
unintegrated gluon distribution and using the $k_T$ factorization
theorem, and we
mentioned some of the limitations of the reduction of the BFKL
equation to 
collinear form.  Here we simply state some of the points to
consider.  In the
\lq unintegrated' formalism it is straightforward to identify the
perturbative
contributions which contribute at {\it all} $Q^2$ and so to avoid
subsuming them
in the input distributions.  Second, there is a natural way to
introduce the
running of $\alpha_S$ in the BFKL formalism, that has increasing
theoretical
support, which for sufficiently small $x$ goes beyond the
Renormalisation Group
behaviour (and so is difficult to implement in the collinear
factorization
approach).  Thirdly, the kinematic constraint along the BFKL
ladder is easy to 
implement in the \lq unintegrated' formalism.  Another point is
that the BFKL
formalism contains all twists, whereas only the leading twist is
retained in the
collinear approach.  Last, but not least, the $k_T$ factorization
approach,
which we may symbolically write as
$F_2 \: = \: F_2^{\gamma g} \otimes f$, is easier to implement. 
The BFKL kernel
and the off-shell gluon structure function $F_2^{\gamma g}$ are
calculable
perturbatively.  We simply use leading order in $\alpha_S$
expressions.  The
$\ln (1/x)$ summations are implicit in the integration over the
{\it entire}
$k_T^2$ phase space of the gluon ladder and in the $k_T$
factorization integrals.

We solved numerically the coupled integral equations for $f (x,
k_T^2)$ and 
$\Sigma (x, k_T^2)$, and we then used the $k_T$ factorization
theorem to calculate
$F_2$ in terms of a two-parameter input form for the gluon, $xg
(x, k_0^2) \:
= \: N (1 - x)^{\beta}$.  The parameters $N$ and $\beta$ are
determined by a
fit to the available small $x$ data for $F_2$. An excellent
description is
obtained.  The data at the smallest values of $x$ give support
for the presence
of the kinematic constraint, as does the extrapolation of the
gluon to 
describe the WA70 prompt photon data at $x \simeq 0.4$.  Notice
that the rise
of $F_2$ with decreasing $x$ is purely of perturbative origin in
our description,
and that we find a significant \lq BFKL' component.

The fact that we achieve an excellent two-parameter fit of the
small $x$ data
for $F_2$ is not, in itself, remarkable.  Other equally good
phenomenological fits
have been obtained.  What is encouraging is that we have a
theoretically
well-grounded and consistent formalism which, with the minimum of
non-perturbative
input, is able to give a good perturbative description of the
observed structure
of $F_2$.  Moreover the BFKL/GLAP components of $F_2$ are decided
by dynamics.
In this way we have made a determination of the {\it universal}
gluon
distribution, $f (x, k_T^2)$, which can be used, via $k_T$
factorization, to
predict the behaviour of other small $x$ observables.  We showed
the predictions
for $F_2^c$ and $F_L$. \footnote{The codes for calculating
$F_2$ are available upon request from a.m.stasto@durham.ac.uk} \\

{\large \bf Acknowledgements}

We thank R.G.\ Roberts and R.S.\ Thorne for valuable discussions.

Two of us (JK and AS) thank the
Physics Department and JK thanks also Grey College of the
University of Durham
for their warm hospitality.  This research has been partially
supported by the
Polish State Committee for Scientific Research (KBN) grant NO 2
P03B 231 08 and
the EU under contracts NO CHRX-CT92-0004/CT93-357.

\newpage
\noindent {\large \bf Figure captions} 
\begin{itemize}
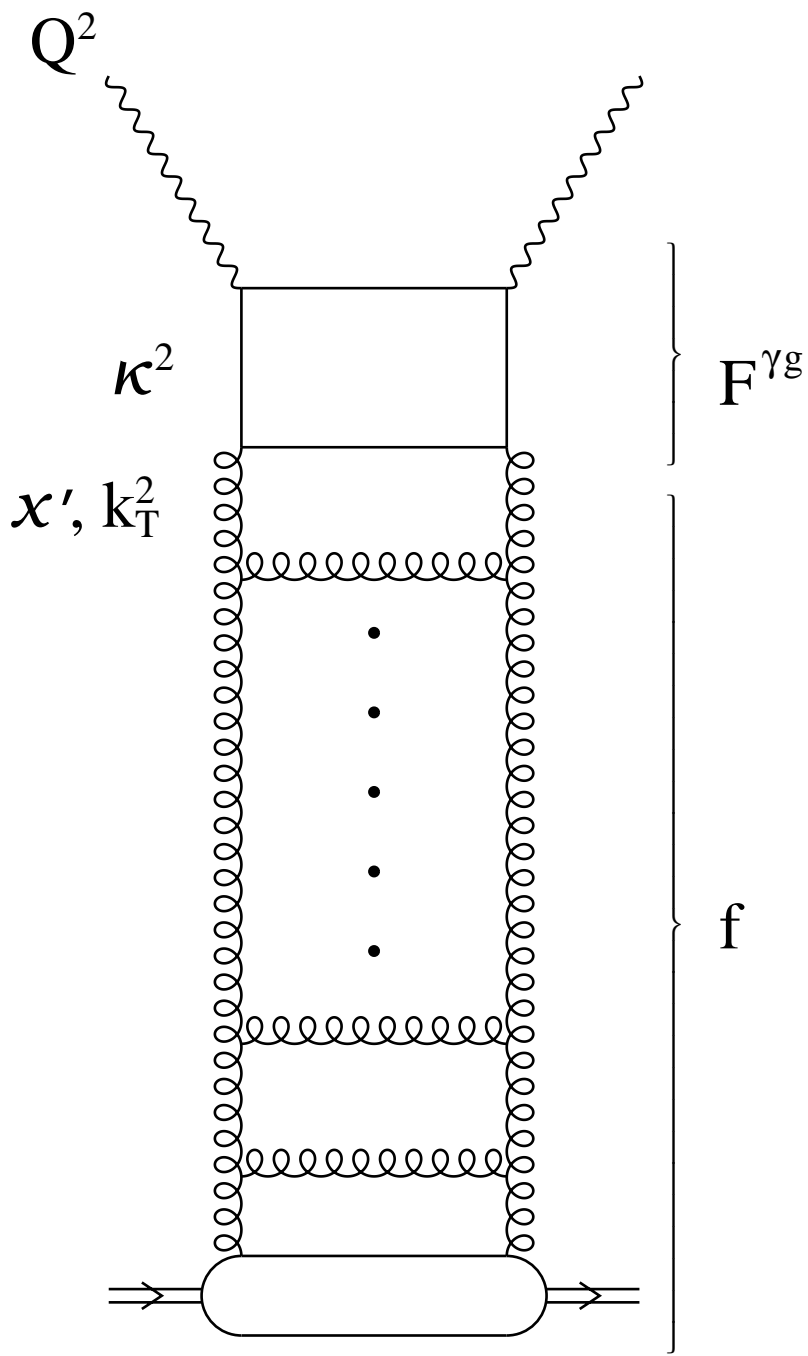
\item[Fig.~1] The diagrammatic representation of the $k_T$
factorization formula
$F_i \; = \; F_i^{\gamma g} \otimes f$.  At lowest order in
$\alpha_S$, the
photon-gluon fusion processes
(or to be precise the structure functions $F_i^{\gamma g}$ of the
virtual gluon)
are given by the quark box shown (together with the crossed box).
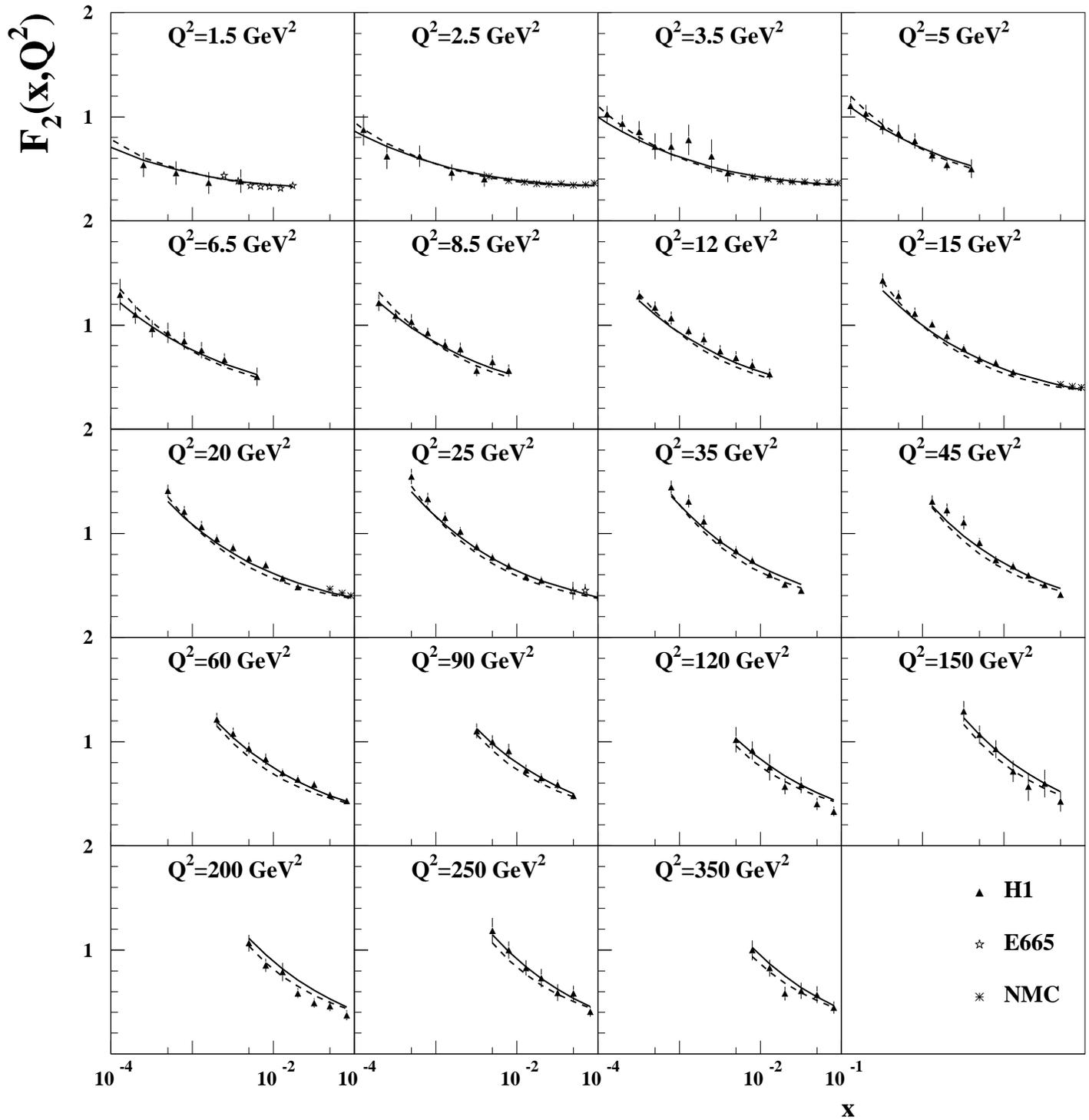
\item[Fig.~2] The two-parameter fit to the $F_2$ data at small
$x$ using
eq. (\ref{eq:z11}) for $f (x, k^2)$ with (continuous curves) and
without
(dashed curves) the kinematic constraint.  The optimum values of
the parameters
$N$ and $\beta$, which describe the input form of the gluon, are
given in Table 
1. The figure shows the H1 data \cite{HERA} together with the
E665 and
NMC measurements \cite{ENB} which occur at the same values of
$Q^2$.
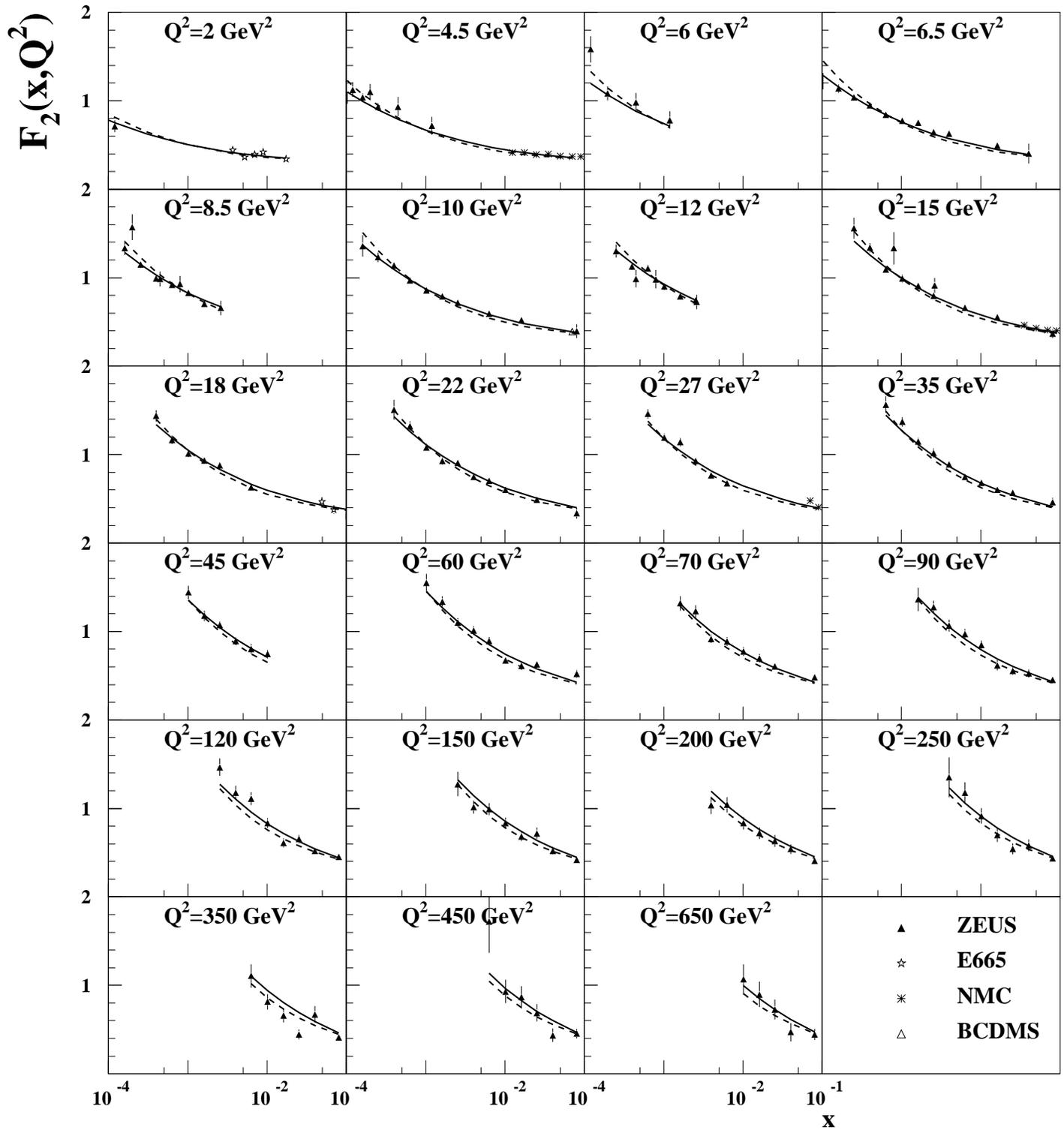
\item[Fig.~3] As for Fig.~1, but for the ZEUS measurements
\cite{HERA} of 
$F_2$, together with the E665, NMC and BCDMS data \cite{ENB}
which occur at the
same values of $Q^2$.
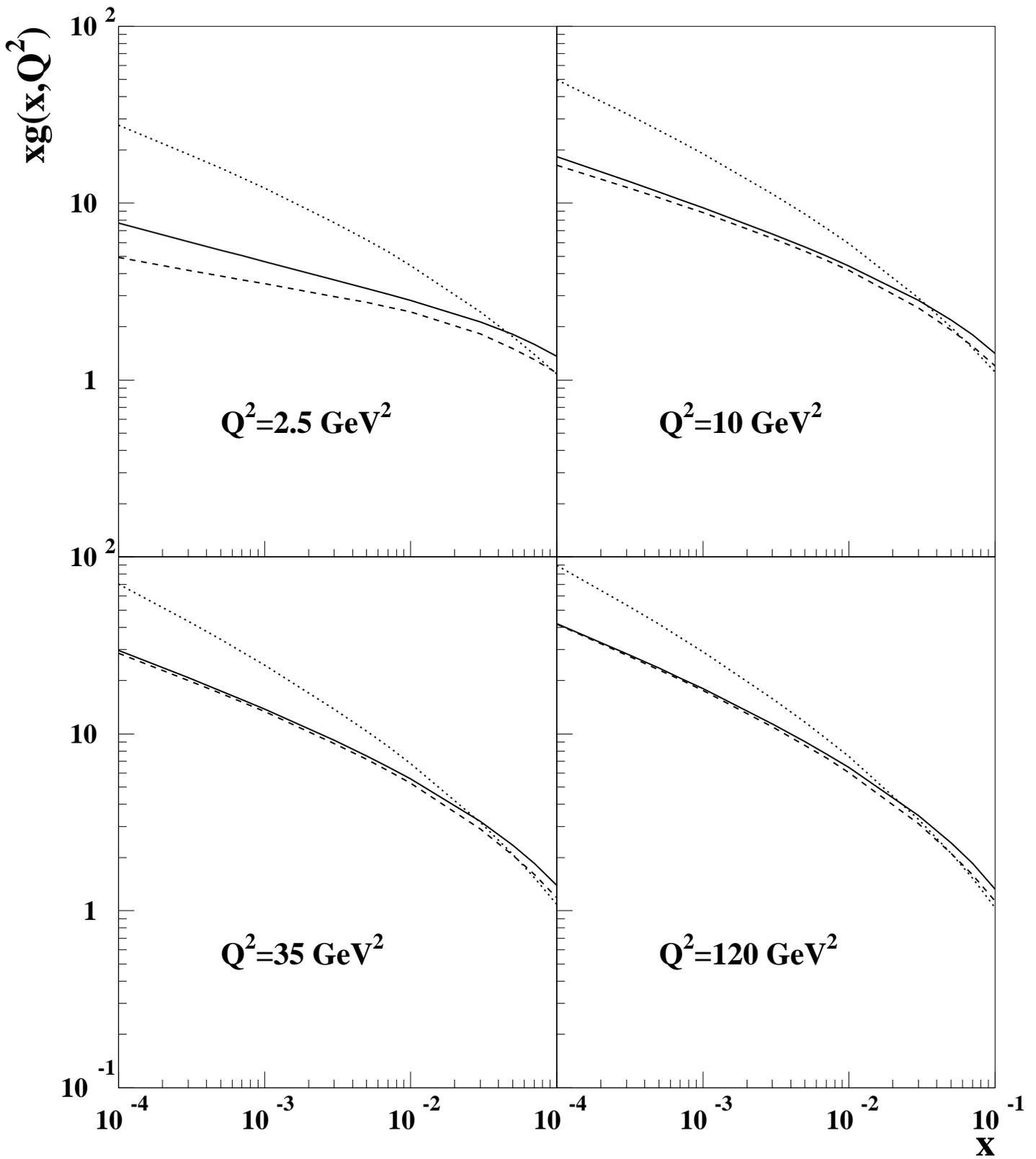
\item[Fig.~4] The continuous curves show the behaviour of the
conventional gluon
distribution $xg (x, Q^2)$ corresponding to fit~1, and calculated
using eq.
(\ref{eq:z1}).  For comparison we also show the gluon
distributions of the MRS (R2)
\cite{MRS} (dashed curve) and GRV \cite{GRV} (dotted curve) 
sets of partons.
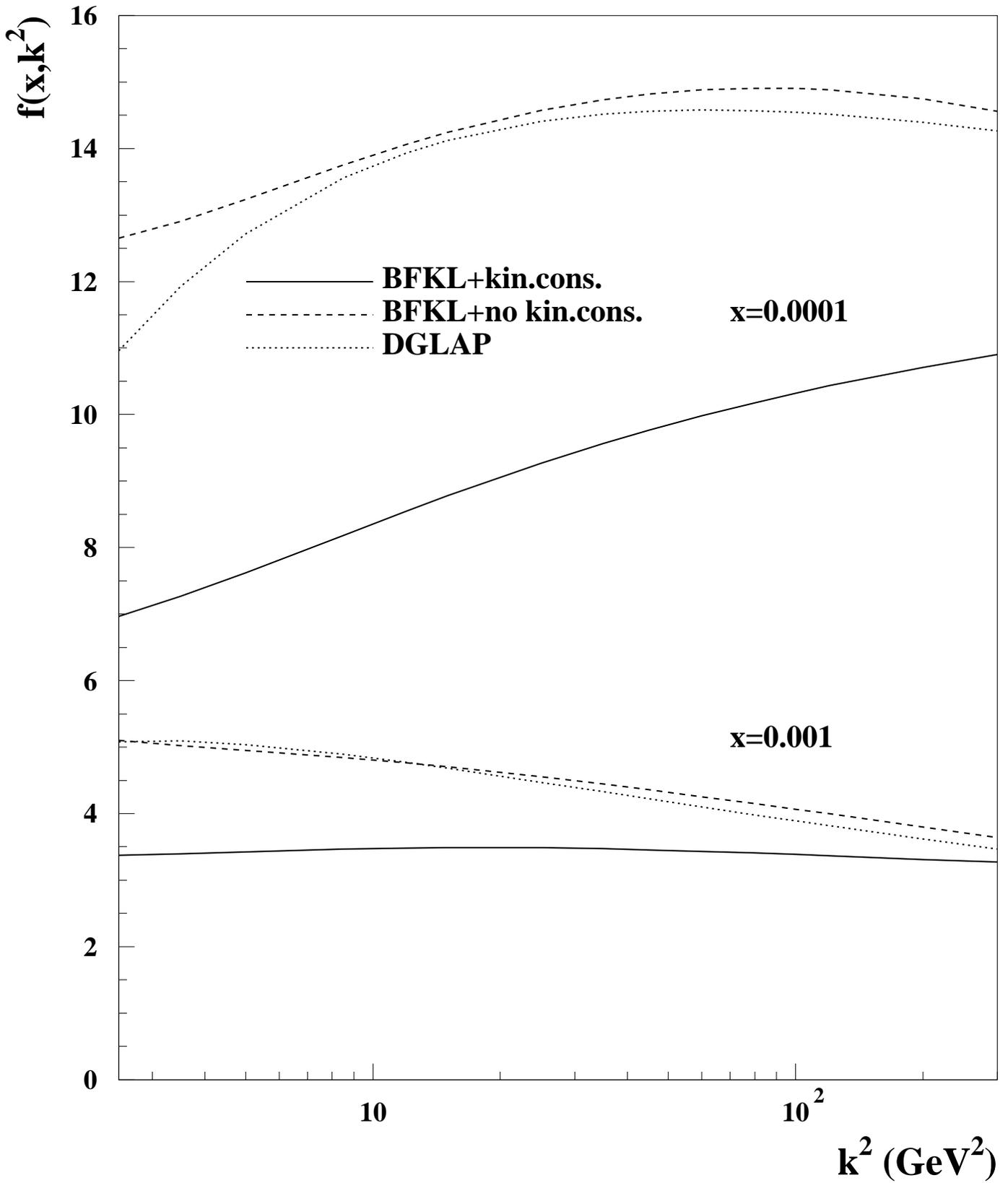
\item[Fig.~5] The unintegrated gluon distribution $f (x, k^2)$ as
a function
of $k^2$ for $x = 10^{-4}$ and $10^{-3}$ obtained by solving the
simultaneous
equations for $f (x, k^2)$ and $\Sigma (x, k^2)$. The continuous
and dashed
curves are obtained by using the unified BFKL/GLAP equation
(\ref{eq:z11}) for
$f (x, k^2)$ with and without the kinematic constraint
respectively.  The dotted
curve corresponds to using GLAP evolution for $f$, eq.
(\ref{eq:q2}).  In each
case the input $xg (x, k_0^2) \: = \: 1.57 (1 - x)^{2.5}$ is
used, where $k_0^2 \:
= \: 1 {\rm GeV}^2$.
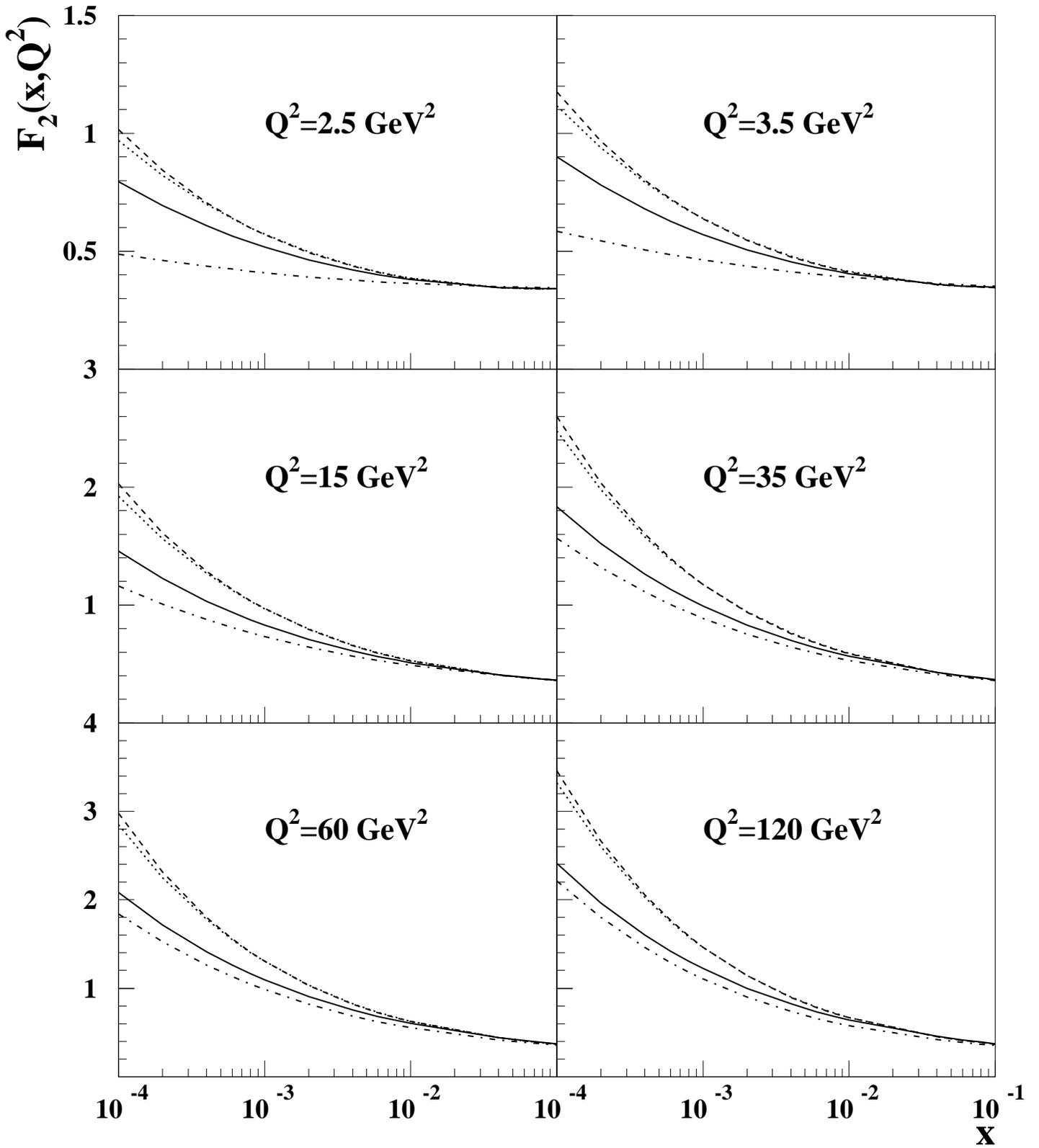
\item[Fig.~6] The light quark contribution to $F_2 (x, Q^2)$ for
various $Q^2$
values obtained from solving different sets of coupled equations
for the gluon
$f$ and the quark singlet $\Sigma$ with, in each case, the input 
$xg (x, k_0^2) \: = \: 1.57 (1 - x)^{2.5}$ where $k_0^2 \: = \: 1
{\rm GeV}^2$.
The continuous and dashed curves come from solving (\ref{eq:z11},
\ref{eq:z13})    
with and without the kinematic constraint.  The dotted curve is
obtained using 
GLAP in the gluon
sector, that is  (\ref{eq:q2},\ref{eq:z13}), whereas the
dot-dashed curve corresponds to pure GLAP
evolution, (\ref{eq:q2},\ref{eq:dglapquark}).
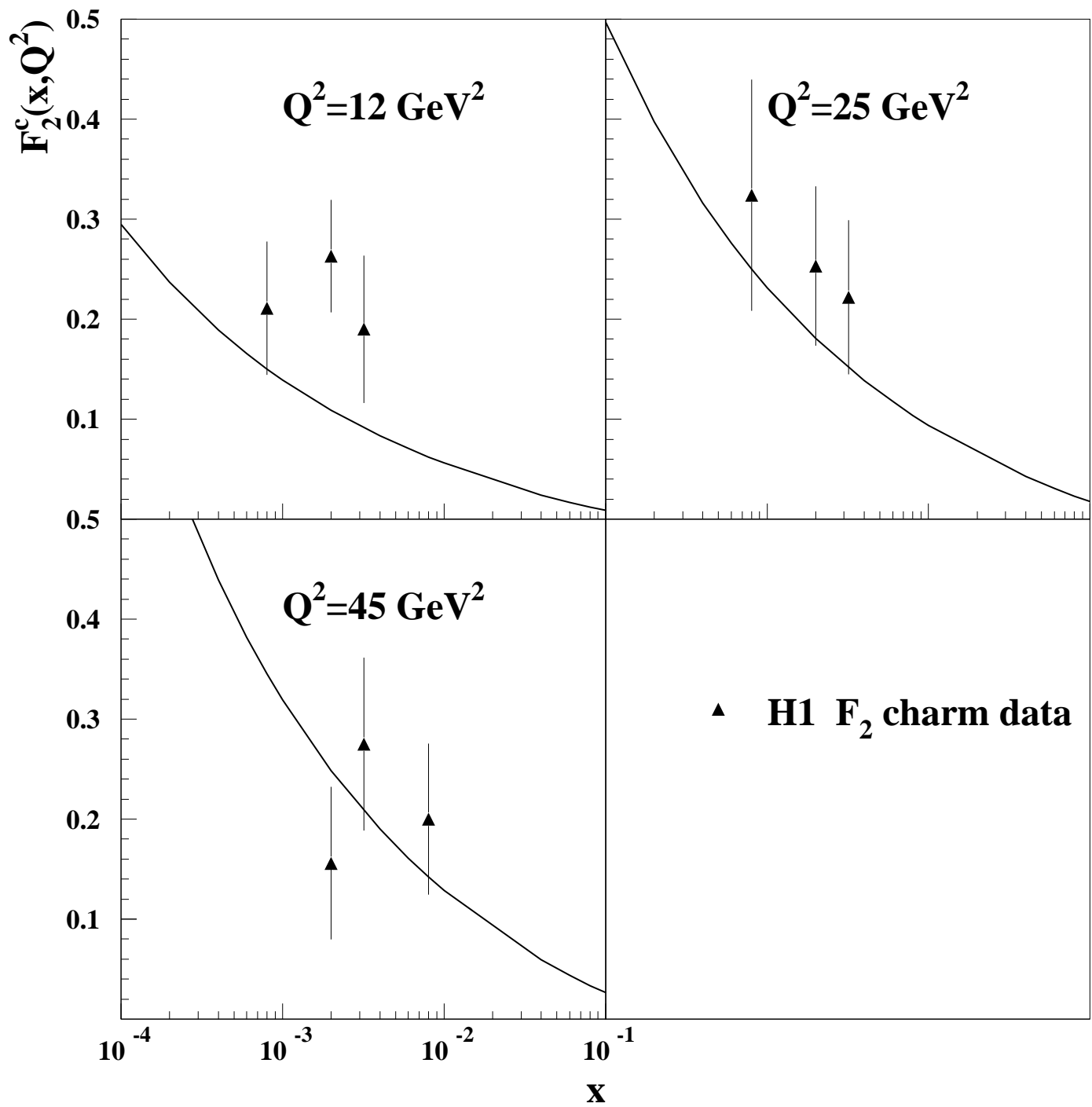
\item[Fig.~7] The predictions for $F_2^c$, compared with H1 charm
data, obtained
from the optimum fit (fit 1).
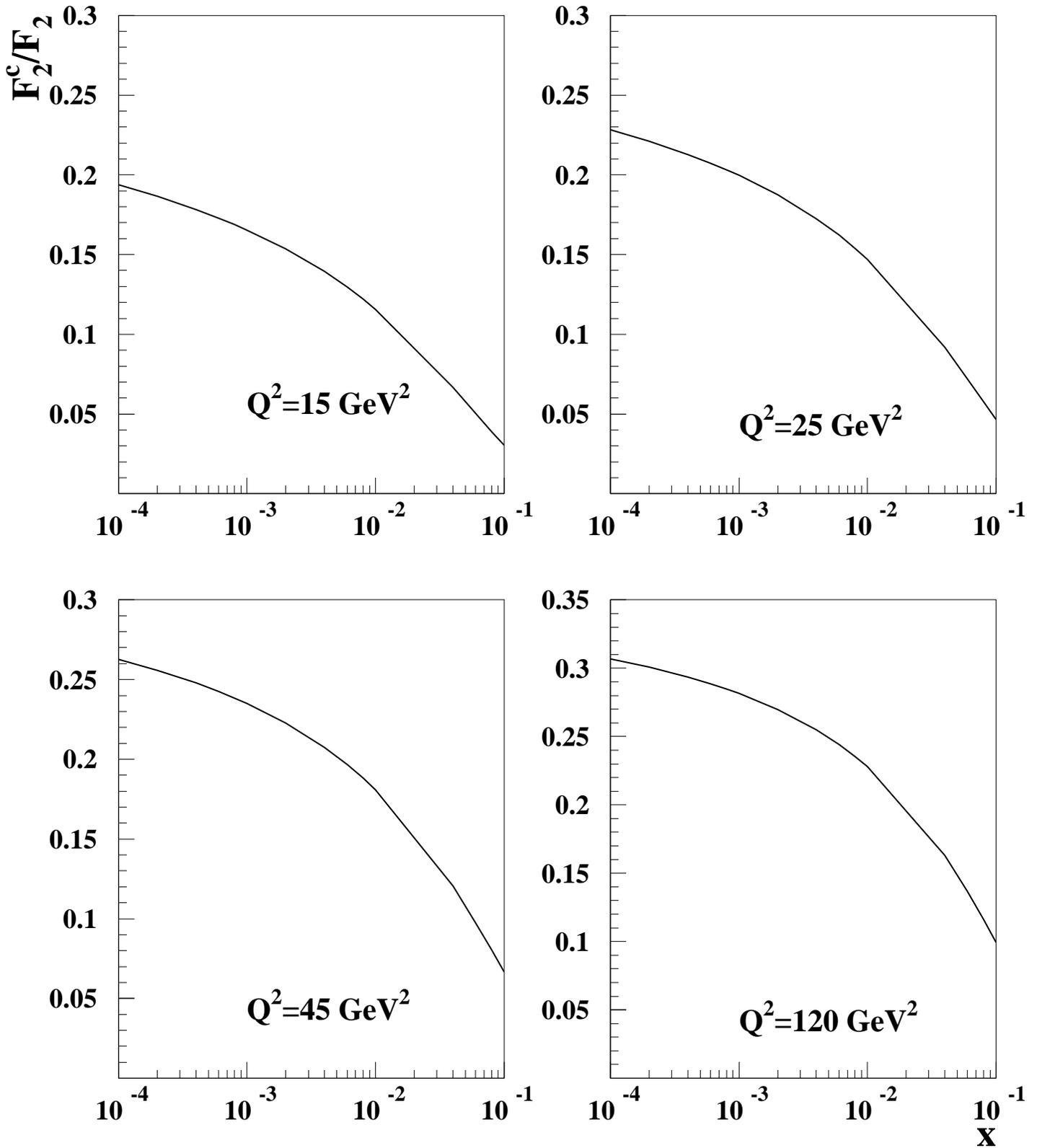
\item[Fig.~8] Ratio $F_2^c/F_2$ for different values of $Q^2$
obtained from fit 1.
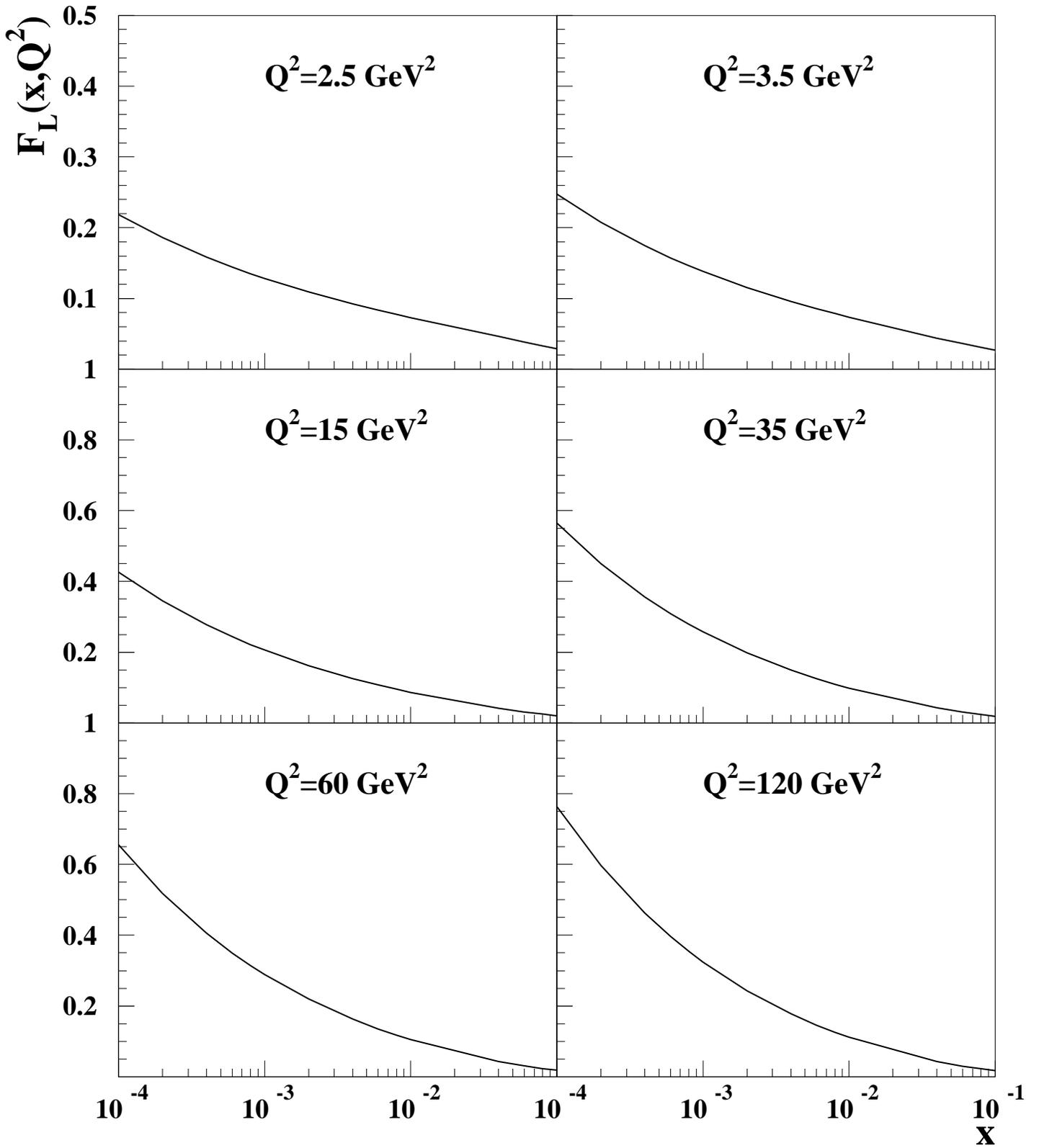
\item[Fig.~9] The prediction for the structure function $F_L$ 
as a function of $x$ for different values of $Q^2$ using the
parameters of fit~1. 
\end{itemize}

\end{document}